\documentclass[useAMS,usenatbib]{mn2e} \usepackage{graphicx}

\title[MAGMO: Carina-Sagittarius]{MAGMO: Coherent magnetic fields in the star forming regions of the Carina-Sagittarius spiral arm tangent}

\author[Green et al.]
       {J. A. Green\thanks{E-mail:james.green@csiro.au}$^{1}$, N. M. McClure-Griffiths$^{1}$, J. L. Caswell$^{1}$, T. Robishaw$^{2}$\newauthor and L. Harvey-Smith$^{1}$ \\ 
$^{1}$CSIRO Astronomy and Space Science, Australia Telescope National Facility, PO Box 76, Epping, NSW 1710, Australia\\
$^{2}$National Research Council, Herzberg Institute of Astrophysics, Dominion Radio Astrophysical Observatory, PO Box 248,\\ Penticton, BC V2A 6J9, Canada}
\date{Accepted 2012 July 13. Received 2012 July 12; in original form 2012 April 18.}

\pagerange{\pageref{firstpage}--\pageref{lastpage}} \pubyear{XXXX}
\begin{document} \maketitle

\label{firstpage}

\begin{abstract}
We present the pilot results of the `MAGMO' project, targeted observations of ground-state hydroxyl masers towards sites of 6.7-GHz methanol maser emission in the Carina-Sagittarius spiral arm tangent, Galactic longitudes 280$^{\circ}$ to 295$^{\circ}$. The `MAGMO' project aims to determine if Galactic magnetic fields can be traced with Zeeman splitting of masers associated with star formation. Pilot observations of 23 sites of methanol maser emission were made, with the detection of ground-state hydroxyl masers towards 11 of these and six additional offset sites. Of these 17 sites, nine are new detections of sites of 1665-MHz maser emission, three of them accompanied by 1667-MHz emission. More than 70\% of the maser features have significant circular polarization, whilst only $\sim$10\% have significant linear polarization (although some features with up to 100\% linear polarization are found). We find 11 Zeeman pairs across six sites of high-mass star formation with implied magnetic field strengths between --1.5 mG and +3.8 mG and a median field strength of +1.6 mG. Our measurements of Zeeman splitting imply that a coherent field orientation is experienced by the maser sites across a distance of 5.3$\pm$2.0\,kpc within the Carina-Sagittarius spiral arm tangent.
\end{abstract}


\section{Introduction}
Masers of the paramagnetic hydroxyl (OH) molecule are clear tracers of magnetic fields. They have a large Zeeman splitting factor (the 1665-MHz transition has the largest known, \citealt{heiles93}) combined with typically narrow linewidths producing fully resolved Zeeman pairs. Unlike molecular absorption in the interstellar medium where Zeeman pairs are blended and only a line-of-sight field component can be determined, the spectrally resolved OH maser Zeeman pairs yield the total magnetic field strength (typically of the order of 1$-$10 mG) independent of field direction. They also provide the orientation of the field along the line of sight, either towards or away from us.

OH masers can be found towards a variety of astrophysical objects, but those which are associated with 6.7-GHz methanol masers exclusively trace regions of high-mass star formation \citep{minier03,xu08}. The recent Methanol Multibeam survey \citep{green09a} has provided a Galaxy-wide census of 6.7-GHz methanol masers, the largest and most complete sample of this maser transition. Through targeted observations of these masers we can study the in situ magnetic fields of high-mass star formation regions spread throughout the Galaxy.

Whilst there has been extensive work examining Galactic magnetic fields using rotation measures towards both extragalactic sources and pulsars \citep[e.g.][]{han06,brown07}, the focus of the current project is examining the magnetic fields traced by Zeeman splitting measurements in regions of high-mass star formation.
A number of studies have been made of the polarized properties of OH masers in regions of star formation, including the recent work of \citet{fish05} and \citet{szymczak09}. 
Observations demonstrate that magnetic fields have generally coherent field direction and magnitude, from  the highest resolution observations with Very Long Baseline Interferometry (VLBI) to the lowest resolution single-dish studies \citep[e.g.][]{fish03b,vlemmings07a}. The consistent fields are observed across scales from tens of parsecs up to several kiloparsecs \citep{reid80,fish02,fish06a}. Given this range of scales, there is an implication that the magnetic fields traced by the masers are tied to the large-scale Galactic magnetic fields, such as those present in the diffuse medium as seen through rotation measures (for example the work of \citealt{brown07} and \citealt{vaneck11} using data such as that of \citealt{taylor93b,han99,mcclure05,haverkorn06,han06}). Several authors have investigated the concept of maser Zeeman splitting tracing the Galactic magnetic field \citep[e.g.][]{davies74,reid90,fish03b,han07}, but mostly with samples of masers collated from a range of heterogeneous observations; the largest set of systematic observations were those of \citet{fish03b}, but these were limited to only 40 star forming regions, all visible from the northern hemisphere, and with only a few masers per spiral arm. 

We have therefore embarked on a homogenous Galaxy-wide study of the polarization properties of OH masers associated with high-mass star formation (the project to study the Magnetic fields of the Milky Way through OH masers, the `MAGMO' project). Targeted observations of OH masers towards sites of 6.7-GHz methanol masers with the Australia Telescope Compact Array (ATCA) allow efficient analysis of Zeeman splitting. Although the ATCA (with a maximum baseline of 6 km) does not have the angular resolution of typical VLBI measurements, many Zeeman patterns recognised at low spatial resolution (e.g. in single-dish spectra) have been positionally confirmed by VLBI \citep[e.g.][and references therein]{fish06a,caswell11vlbi1}.
Thus the observations  provide the opportunity to determine whether the orientation of large-scale magnetic fields in the diffuse medium can be conserved in the contraction to the high densities of star formation (despite the turbulence and compression inherent in these processes). 

The pilot results presented here cover sources between Galactic longitudes 280$^{\circ}$ and 295$^{\circ}$, the direction in which the Carina-Sagittarius arm is tangential to the line of sight. The Methanol Multibeam survey identified 23 sites of 6.7-GHz methanol masers \citep{green12mmb4} and eight sites of OH maser emission are known to exist from previous observations \citep{caswell87b,caswell98,caswell04a}. The paper is structured such that section 2 describes the observing procedure together with the strategy for maser and Zeeman pair identification; section 3 presents the detections of the survey and properties of the maser sites; and section 4 discusses the results in the context of Galactic magnetic fields.

\begin{figure}
\begin{center}
\renewcommand{\baselinestretch}{1.1}
\includegraphics[width=8cm]{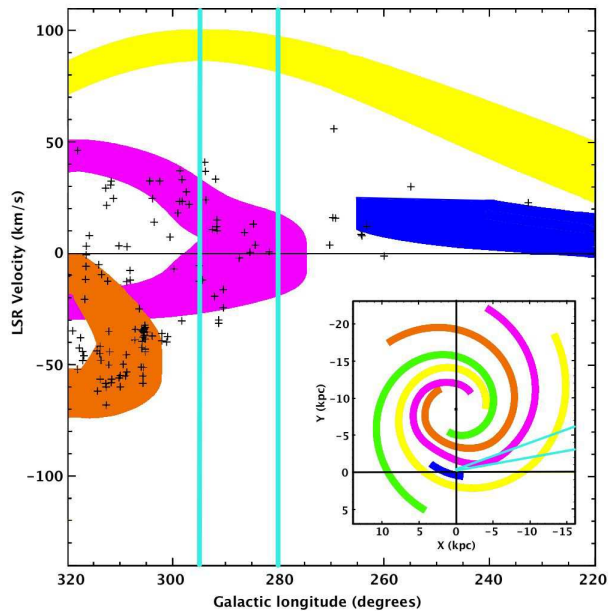}  
\caption{\small Galactic longitude versus velocity (LSR) distribution of 6.7-GHz methanol maser sources (`plus' symbols) from the Methanol Multibeam survey \citep{green12mmb4}, in the region 320$^{\circ}$ to 220$^{\circ}$.  Inset image shows the top-down view of the spiral arms within the Galaxy \citep{taylor93a} where yellow loci represent the Perseus spiral arm; purple - Carina-Sagittarius; orange - Crux-Scutum; green - Norma;  blue -Local arm 
(Orion-Cygnus).  Coloured loci in the main diagram show the same spiral arms transferred to the longitude-velocity domain via a flat rotation curve with circular rotation of 246\,km\,s$^{-1}$ \citep{reid09a, bovy09};  their thickness incorporates an arm width of 1\,kpc and a velocity tolerance of $\pm$7\,km\,s$^{-1}$. Cyan lines in both the main figure and inset image delineate the longitude region studied in this paper, 280$^{\circ}$ to 295$^{\circ}$.}
\label{LVdiagram}
\end{center}
\end{figure}

\begin{table*} \centering \caption{\small 6.7-GHz methanol maser targets from \citet[][references therein]{green12mmb4}. The target velocity is the approximate midpoint of the range over which methanol emission is seen. Superscript numbers refer to pairs of sources observed with the same OH observation. $\dagger$ Distances are kinematic distances calculated using a flat rotation curve with $\theta$ of 246\,km\,s$^{-1}$ and R$_{\odot}$ of 8.4\,kpc \citep{reid09a, bovy09}. The midpoint LSR velocity of the maser is used after it has been corrected for the best estimates of the solar motion: $U_{\odot}$ = 11.1\,km\,s$^{-1}$, $V_{\odot}$ = 12.2\,km\,s$^{-1}$, $W_{\odot}$ = 7.25\,km\,s$^{-1}$ \citep{reid09a,mcmillan10,schonrich10}. Where appropriate, near/far ambiguity resolutions are from \citet{green11b}.} 
\begin{tabular}{lcrrr}
\hline
\multicolumn{1}{c}{Source Name} & \multicolumn{2}{c}{Equatorial Coordinates} & \multicolumn{1}{c}{Velocity} & \multicolumn{1}{c}{Distance}\\
\ (~~~$l$,~~~~~~~$b$~~~)    &       RA(J2000)        &       Dec(J2000)       & \multicolumn{1}{c}{$\rm V_{\rm target}$} &\multicolumn{1}{c}{Estimate$\dagger$}\\
\ (~~~$^\circ$~~~~~~~$^\circ$~~~) & (h~~m~~~s) & (~$^\circ$~~ $'$~~~~$''$) & \multicolumn{1}{r}{(km\,s$^{-1}$ )}& \multicolumn{1}{c}{(kpc)}  \\
\hline
281.710$-$1.104&10 05 05.63& $-$56 57 24.7&2.0 &4.2$\pm$0.8\\
284.352$-$0.419&10 24 10.89& $-$57 52 38.8&7.0& 5.4$\pm$0.8\\
284.694$-$0.361&10 26 36.29& $-$58 00 34.3&13.0& 6.0$\pm$0.7\\
285.337$-$0.002&10 32 09.62& $-$58 02 04.6&$-$3.0& 4.5$\pm$0.8\\
286.383$-$1.834&10 31 55.12& $-$60 08 38.6&9.0& 6.0$\pm$0.7\\
287.371+0.644&10 48 04.44& $-$58 27 01.0&$-$2.0& 5.2$\pm$0.7\\
290.374+1.661&11 12 18.10& $-$58 46 21.5&$-$25.0& 2.9$\pm$1.9\\
290.411$-$2.915&10 57 33.89& $-$62 59 03.5&$-$16.0&4.1$\pm$1.1\\
291.270$-$0.719$^{1}$&11 11 49.44& $-$61 18 51.9&$-$26.0&3.1$\pm$1.9\\
291.274$-$0.709$^{1}$&11 11 53.35& $-$61 18 23.7&$-$29.0&3.1$\pm$1.9\\
291.579$-$0.431$^{2}$&11 15 05.76& $-$61 09 40.8&15.0&7.4$\pm$0.6\\
291.582$-$0.435$^{2}$&11 15 06.61& $-$61 09 58.3&9.5&7.4$\pm$0.6\\
291.642$-$0.546&11 15 14.32& $-$61 17 26.7&12.0&7.6$\pm$0.6\\
291.879$-$0.810&11 16 17.35& $-$61 37 20.7&33.0&9.4$\pm$0.6\\
292.074$-$1.131&11 16 51.24& $-$61 59 32.6&$-$19.0&4.2$\pm$1.1\\
292.468+0.168&11 23 42.17& $-$60 54 33.5&16.0&8.1$\pm$0.6\\
293.723$-$1.742&11 28 32.97& $-$63 07 18.6&24.5&9.1$\pm$0.6\\
293.827$-$0.746&11 32 05.56& $-$62 12 25.3&37.0&10.2$\pm$0.6\\
293.942$-$0.874&11 32 42.09& $-$62 21 47.5&39.0&10.5$\pm$0.6\\
294.337$-$1.706&11 33 49.91& $-$63 16 32.5&$-$12.0&6.1$\pm$0.7\\
294.511$-$1.621&11 35 32.25& $-$63 14 43.2&$-$9.0&6.4$\pm$0.7\\
294.977$-$1.734$^{3}$&11 39 13.94& $-$63 29 04.6&$-$6.0&6.8$\pm$0.7\\
294.990$-$1.719$^{3}$&11 39 22.88& $-$63 28 26.4&$-$12.0&6.3$\pm$0.8\\
\hline
\end{tabular} 
\label{targetstable}
\end{table*}

\section{Observations}
Using the ATCA we observed all four transitions of ground-state OH maser emission (1612.2310, 1665.4018, 1667.359 and 1720.530 MHz) towards all known 6.7-GHz methanol maser sites between longitudes 280$^{\circ}$ and  295$^{\circ}$ \citep[][and references therein]{green12mmb4}. Target sources are listed in Table \ref{targetstable} and the region in longitude-velocity space is shown in Figure\,\ref{LVdiagram}. We adopted the CFB 1M-0.5k mode of the Compact Array Broadband Backend \citep{wilson11} with three concatenated `zoom' bands at each of the maser transition frequencies (two transitions observed at each intermediate frequency). This correlator configuration provided 2048 channels over 1 MHz giving 0.5 kHz channel spacings (velocity channel separations of 0.091, 0.088, 0.088 and 0.085\,km\,s$^{-1}$ for the four respective transitions), with all four Stokes polarization products. Concatenated zooms were centred at 1612.0, 1665.5, 1667.5 and 1720.5 MHz with respective velocity coverages of  372, 360, 360 and 348\,km\,s$^{-1}$. Observations were made with the 6-km array (configuration 6B) and for each source consisted of five `cuts' of six minutes spread over a 12 hour period. ATCA calibrator B1049$-$534 (J1051$-$5344) was observed as a phase calibrator (for three minutes every $\sim$18 minutes) and PKS B0823$-$500 for primary flux calibration. Observations were obtained  2011 September 11--13. 

The data were reduced and processed with the {\sc miriad} software package using standard techniques \citep{sault04}. The primary beam has a full width at half maximum (FWHM) of $\sim$28$'$. The synthesized beam has a FWHM in right ascension of $\sim$7$''$ and a FWHM in declination of  7$''$ to 8$''$ (dependent on declination). The rms position errors of the observations are estimated to be $\sim$0.4$''$ in each coordinate. This estimate is based on the phase calibrator being typically offset by 10$^{\circ}$ from the beam centre of the target pointing, and typical signal-to-noise ratios (see \citealt{caswell98} for further details). The channel rms noise ($\sigma$) was $\sim$50\,mJy.  Masers were detected as three or more channels with positionally coincident emission above 3$\sigma$. Zeeman pairs were identified through coincident  right hand circularly polarized (RHCP) and left hand circularly polarized (LHCP) features (to within 0.4$''$), each feature found through fitting two-dimensional Gaussians to individual channel maps (`centroid positions').

\section{Results}\label{resultssection}
We detected OH emission towards 11 of the 23 methanol maser targets, coincident to within the positional errors, and six additional sites (two offset by $\sim$5$''$ and $\sim$10$''$, the remainder offset by 2$'$ to 23$'$). In total there were 17 detections, nine of which are new discoveries. Detections are listed in Table \ref{resotable} and the Stokes ($I$,$Q$,$U$,$V$) spectra are shown in Figure \ref{spectra}. The RHCP and LHCP features, derived from RHCP = $(I+V)/2$ and LHCP = $(I-V)/2$,  are shown in Figure\,\ref{RLspectra}. Although each methanol target site was observed at the frequencies of all four ground-state OH transitions, only the detections are shown in Figures \ref{spectra} and \ref{RLspectra}. Six sources are found offset from the field centre, and primary beam correction factors (only one of which exceeds 1.1) are listed in Table \ref{resotable}, but not applied to the spectra shown in the Figures. 

None of the methanol target sources, nor the six additional sources, were found to exhibit 1720-MHz emission; 290.374+1.661 was found by \citet{caswell04a} to have 1720-MHz at the nearby site 290.375+1.666, but this source is no longer present at our detection limit ($<$0.20 Jy). From all the detections (methanol and non-methanol associated) only one source was found to exhibit 1612 MHz emission, 284.351$-$0.418. This is a new detection of this transition with accompanying 1665 and 1667 MHz OH but no directly associated methanol. The rarity of these satellite transitions is in accordance with \citet{caswell99}. Of the 17 OH maser sites, nine are 1665-MHz only, and eight are 1665-MHz and 1667-MHz combined. 

Sidelobe responses have been labelled in the spectra, but here we give some further details: the features seen in the spectra of 291.579$-$0.431 between 17\,km\,s$^{-1}$ and 20\,km\,s$^{-1}$ are a sidelobe of 291.610$-$0.529; 291.579$-$0.434 contains a sidelobe of 291.579$-$0.431 between 11\,km\,s$^{-1}$ and 14\,km\,s$^{-1}$ and of 291.610$-$0.529 between 17\,km\,s$^{-1}$ and 18.5\,km\,s$^{-1}$. 

\subsection{Polarization properties}
The full breakdown of the polarization properties is given in Table\,\ref{poltable}. We find 80 maser features in the current study, of which nine have statistically significant linear polarization and 57 have statistically significant circular polarization. We see features with fractional linear polarization between 22\% and 95\%. The highest linearly polarized features are found in the 1665-MHz emission from 291.879$-$0.810 and 294.511$-$1.621. We see features with fractional circular polarization between 6\% and 100\%, with 19 features consistent with 100\% circular polarization. 

\subsection{Magnetic field measurements}
We found 11 Zeeman pairs across five sites of star formation, with the RHCP and LHCP components coincident to within 0.4$''$ of the fitted centroid positions. 
To derive values of the magnetic field, we have used splitting factors for the separation of LHCP and RHCP of 0.123, 0.590, 0.354, 0.113 and 0.0564 km\,s$^{-1}$ per mG for the transitions at 1612, 1665, 1667, 1720 and 6035 MHz respectively and positive sign denoting RHCP at larger velocity than LHCP.  For the ground-state transitions, these are essentially the values as given by \citet{palmer67}, and with more detailed derivation by \citet{cook75}, and at 6035 MHz as given by \citet{yen69}. For the ground-state satellite transitions at 1612 and 1720 MHz, the splitting factors correspond to the pair of 
magnetic sub-components of smallest splitting, rather than blends with the weaker sub-components of larger splitting that are theoretically present. This assumption has been justified by \citet{caswell04a}, \citet{wright04b} and others, based on the narrowness of features seen in well separated pairs, and from theoretical considerations \citep{gray92} suggesting that these components will be especially dominant in the masing situation. The resulting magnetic fields from our measurements are listed in Table\,\ref{RLtable} together with three additional measurements from the literature (for 285.263$-$0.050, 290.375$-$0.529 and 294.511$-$1.621). Collectively we find three negative field measurements and 11 positive field measurements with the field strengths for our measurements between $-$1.5 mG and +3.8 mG and those of the literature between +1.2 mG and +8.9 mG. The field measurements are classified as `A' where there is an unambiguous  Zeeman pair associated to within the positional errors; and `B' where there are more than one RHCP or LHCP component that can be associated within the positional errors, or the positional coincidence is weaker due to a low signal-to-noise ratio. We also note that the errors on the magnetic field strength are based on the error in velocity (half the width of a channel for each component).

\subsubsection{Comparison with previous measurements}
There are only two sources (285.263$-$0.050 and 291.610$-$0.529) for which the previously published spectra have high spectral resolution, good sensitivity and clear Zeeman pair identification.  
The previously published spectra of 291.610$-$0.529, at both 1665 and 1667 MHz \citep{caswell87a}, show remarkable similarity with the present data  obtained more than 20 years later. The Zeeman patterns inferred from our data can also be identified in the earlier noisier spectra.
285.263$-$0.050 has shown much greater variability not only at 1665 MHz, but also at 6035 MHz. Measurements of the 6035-MHz transition of 285.263$-$0.050 taken in 1994 indicated a magnetic field of nearly 10 mG, inferred from multiple features \citep{caswell95c}. However, seven years later, all emission had faded by an order of magnitude, with only one Zeeman pair remaining visible \citep{caswell03}.  At the 1665 MHz transition, the spectrum in 1993 (shown by \citealt{caswell95c}) was best interpreted as the combination of six Zeeman pairs, with the strongest field +9.1 mG and the weakest +2.7 mG (and a median field of +8.0 mG).  Our present spectrum, although still showing strong emission, has changed dramatically with the loss of both features of three pairs, and the RHCP and LHCP components respectively lost from two others;  the sole recognisable Zeeman pair now seen is the same as the 1993 pair revealing a field of +2.7 mG. Our 1667-MHz spectrum corroborates this, displaying a Zeeman pair with similar mean velocity and a field of +3.2 mG.  

\subsection{Characterisation of sources by maser transition, with evolutionary implications}
The combination of maser transitions observed towards a site of high-mass star formation can be used to characterise the evolutionary state of the region \citep[e.g.][]{ellingsen07,breen10a}. Approximately 80\% of star forming OH maser sites are believed to have associated methanol \citep{caswell98} and we find 65\% (11 of the 17) of the OH sources are directly associated (to within the positional errors) with methanol. With a Poisson statistical error of $\pm$18\% this is a result consistent with expectations. Sites without methanol are believed to have a more evolved Ultra-Compact H{\sc ii} region which has led to the quenching of the methanol maser emission. Thus we expect the six sites without methanol (see Table \ref{resotable}) to be the more evolved.

The proportion of methanol maser sites within our Galaxy which have associated OH maser emission is not well determined. This statistic indicates the proportion of high-mass star forming regions traced by 6.7-GHz methanol emission with a more developed region of ionized hydrogen (those with OH emission). We find 48\% (11 of the 23) of the methanol sources are associated with OH maser emission, implying approximately half of the methanol sources are in a more evolved stage of high-mass star formation. With an expected total Galactic population of $\sim$1200 methanol masers \citep{walt05}, we can therefore expect of the order of 600 star forming OH masers in the Galaxy. 

Five of the 17 sites are found to have associated water maser emission \citep{breen10b} and a further four have possible associated water masers within 30$''$ \citep{walsh11}. Water masers are known to trace outflows from the sites of high-mass star formation, and five of these nine associations are sites that also have 6.7-GHz methanol maser emission. Two of the sites (287.371+0.644 and 293.942$-$0.874) have associated 12.2-GHz methanol maser emission \citep{breen12}. 

The different combinations of maser species thus suggests that the 12 sites without OH detections are the youngest, the 11 sites with both methanol and OH then follow and the six sites with OH only are the oldest. The different combinations of OH maser transitions, ground-state and excited-state, may reveal further evolutionary differences, and this will be examined in a more detailed follow-up of the complete MAGMO survey.

\subsection{Non-detections and absorption}\label{absection}
We did not detect any OH emission in 12 of the 23 target 6.7-GHz methanol sources ($<$0.25 Jy to 5$\sigma$ confidence). However, near two methanol sites there was OH emission which was not coincident, but sufficiently near to be attributed to the same star forming cluster. Specifically, 291.582$-$0.435 does not show coincident OH emission, but its close companion 291.579$-$0.431 (offset by $\sim$20$''$) does, and a further OH maser site, 291.5790$-$0.434, is offset 10$''$ north. Towards 291.642$-$0.546, while no OH emission was coincident, a detection was made at the nearby site  291.610$-$0.529 (offset by $\sim$2$'$). 

\begin{figure}
\begin{center}
\renewcommand{\baselinestretch}{1.1}
\includegraphics[width=8cm]{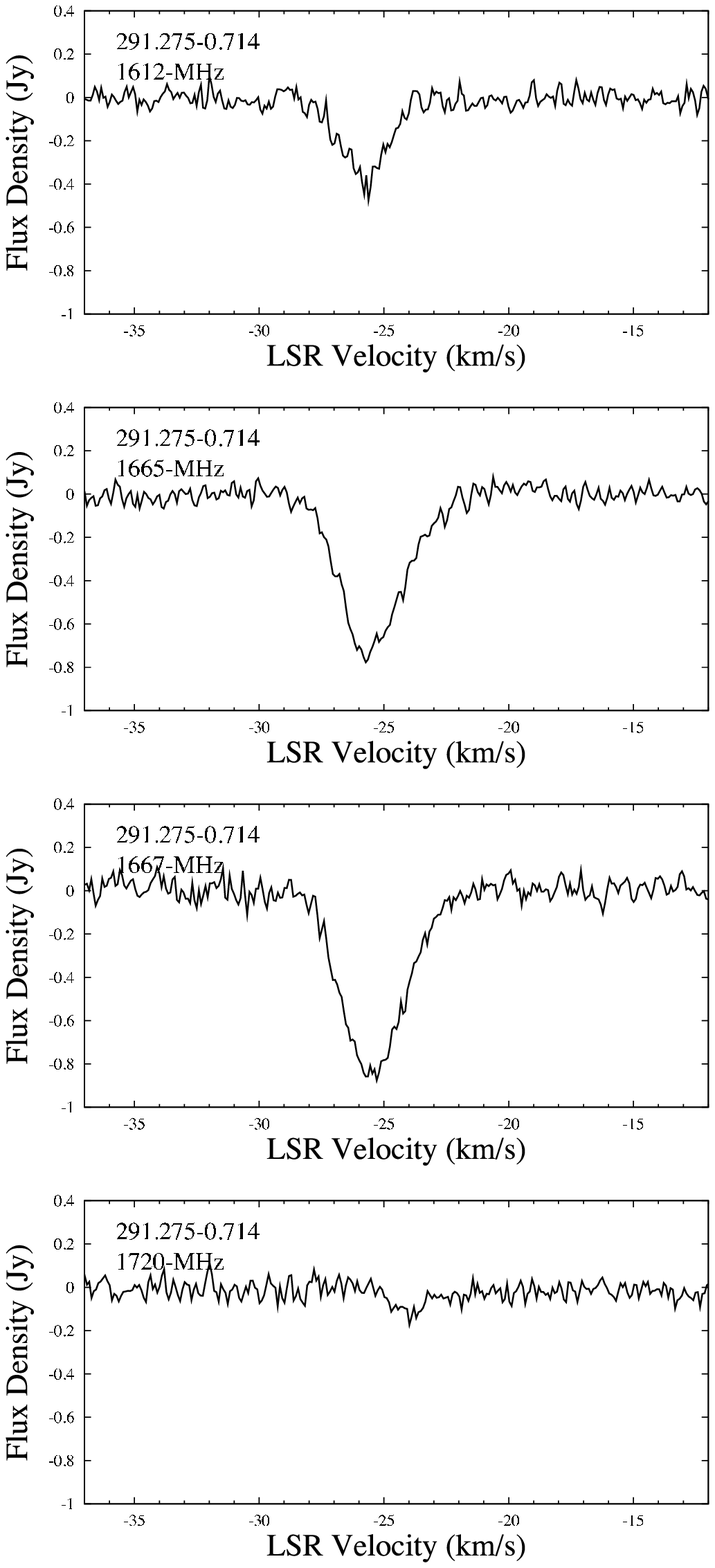}  
\caption{\small OH absorption spectra located at RA(J2000)11$^{\rm h}$11$^{\rm m}$52.73$^{\rm s}$, Dec(J2000)~$-$61$^{\circ}$18$'$44.59$''$ (291.275-0.714), near the continuum peak of NGC 3576 (see section \ref{absection}).}
\label{290p270abs}
\end{center}
\end{figure}

The field containing 291.274$-$0.709 and its close companion 291.270$-$0.719 shows extended absorption. This was found to be centred approximately on RA(J2000) 11$^{\rm h}$11$^{\rm m}$52.73$^{\rm s}$, Dec(J2000) $-$61$^{\circ}$18$'$44.59$''$ (291.275$-$0.714). It is essentially coincident (to within 7$''$) with the poorly defined 3.4-cm extended radio continuum peak of the H{\sc ii} region NGC 3576, mapped with 7$''$ resolution by \citet{depree99}. The separate absorption spectrum centred at this position for each transition is shown in Figure\,\ref{290p270abs}. 

\begin{figure*}
\begin{center}
\renewcommand{\baselinestretch}{1.1}
\includegraphics[width=18cm]{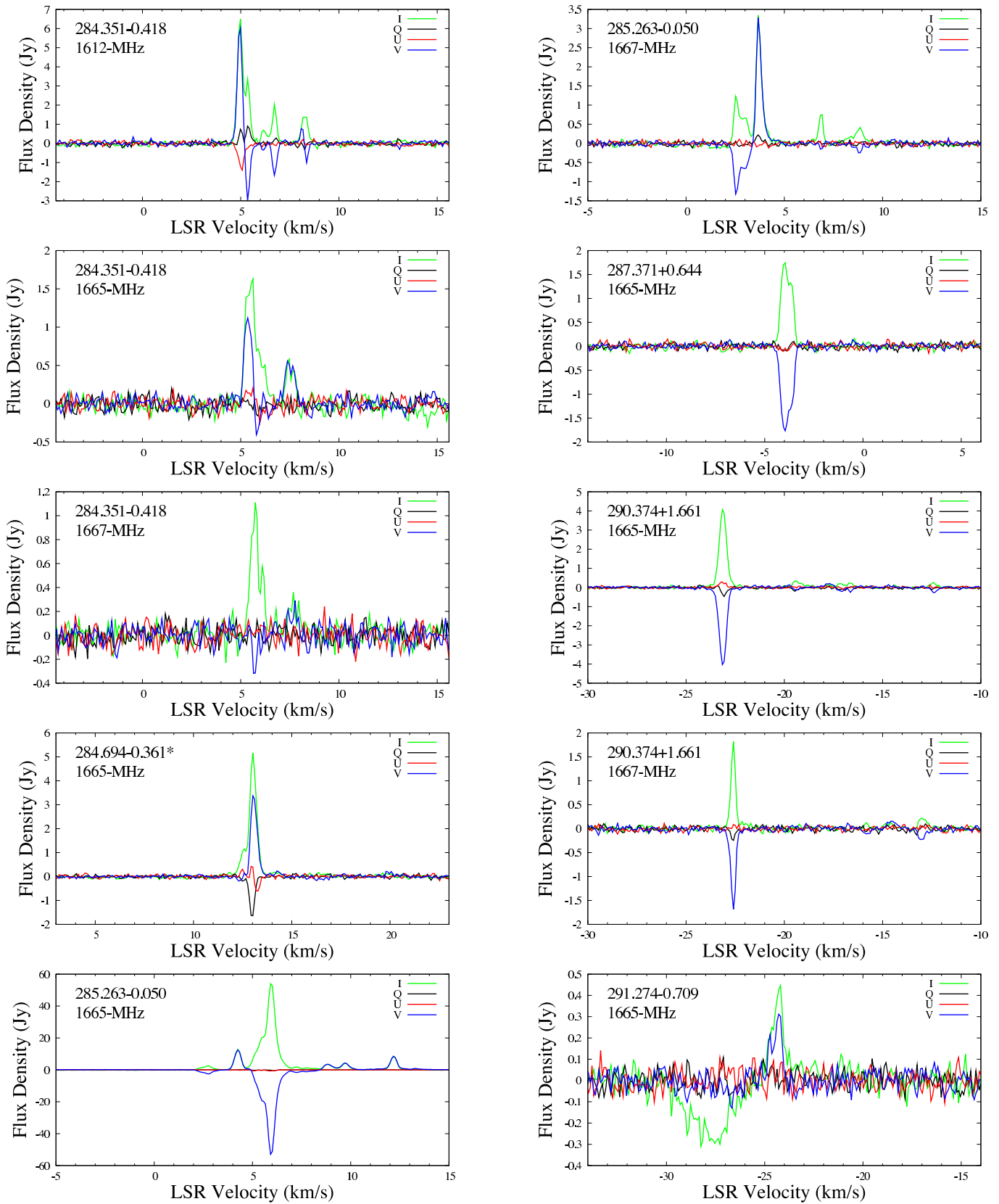}  
\caption{\small Stokes parameters for all  detections of ground-state OH maser emission: green - $I$; black - $Q$; red - $U$; and blue - $V$. * denotes new sources. `S/L' denotes side-lobe response. The flux density of 285.263$-$0.050 should be scaled by a factor of 1.08 to account for the primary beam correction.}
\label{spectra}
\end{center}
\end{figure*}

\begin{figure*}
\begin{center}
\addtocounter{figure}{-1}
\renewcommand{\baselinestretch}{1.1}
\includegraphics[width=18cm]{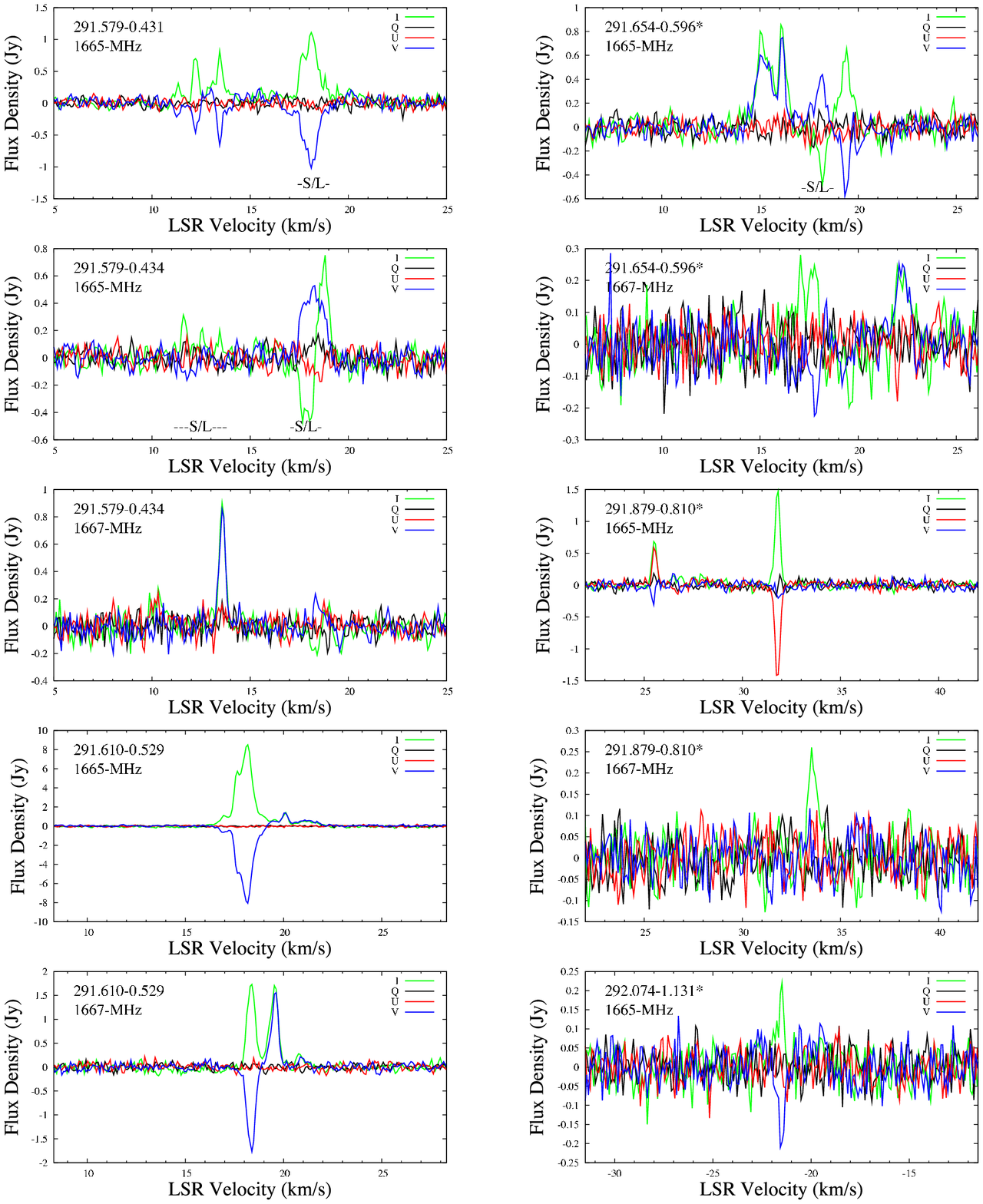}  
\caption{\small cont. The flux densities of 291.610$-$0.529 and 291.654$-$0.596 should be scaled by factors of 1.01 and 1.02 respectively to account for the primary beam correction. }
\label{spectra}
\end{center}
\end{figure*}

\begin{figure*}
\begin{center}
\addtocounter{figure}{-1}
\renewcommand{\baselinestretch}{1.1}
\includegraphics[width=18cm]{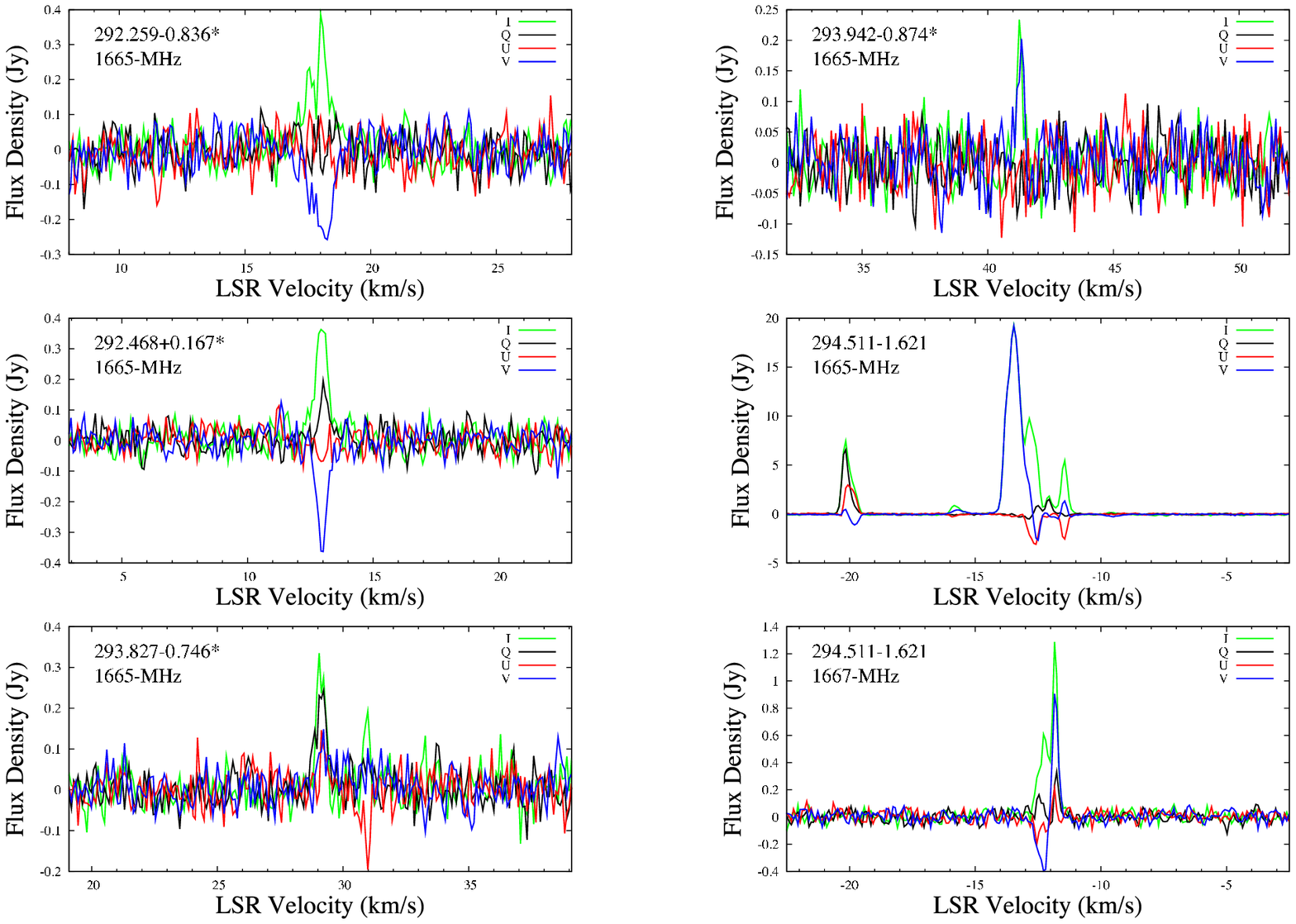}  
\caption{\small cont. The flux density of 292.259$-$0.836 should be scaled by a factor of 6.58 to account for the primary beam correction. }
\label{spectra}
\end{center}
\end{figure*}

\begin{figure*}
\begin{center}
\renewcommand{\baselinestretch}{1.1}
\includegraphics[width=18cm]{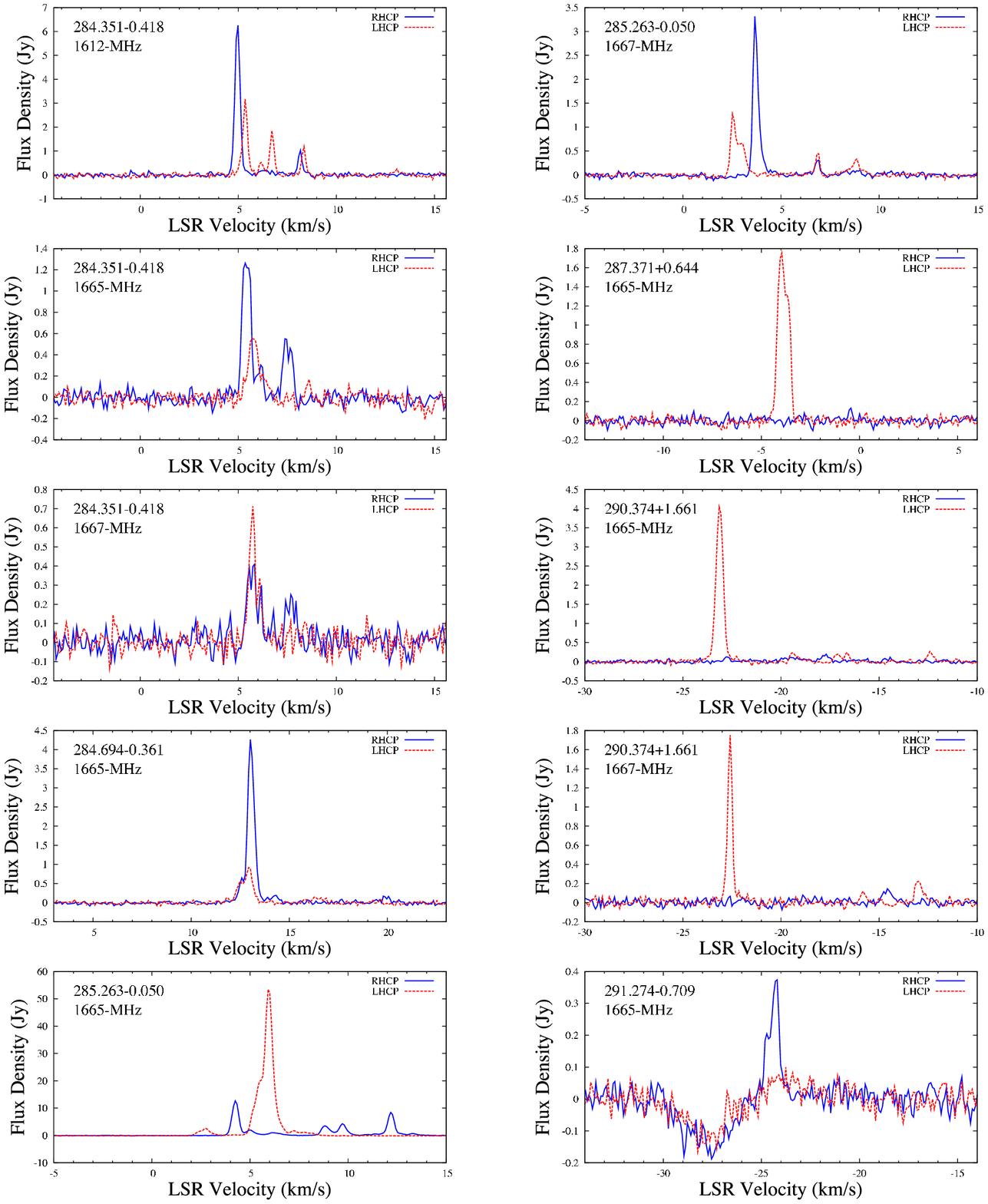}  
\caption{\small Spectra of the right hand circularly polarized (solid blue) and left hand circularly polarized (red dashed) maser features. `S/L' denotes side-lobe response. The flux density of 285.263$-$0.050 should be scaled by a factor of 1.08 to account for the primary beam correction.}
\label{RLspectra}
\end{center}
\end{figure*}

\begin{figure*}
\begin{center}
\addtocounter{figure}{-1}
\renewcommand{\baselinestretch}{1.1}
\includegraphics[width=18cm]{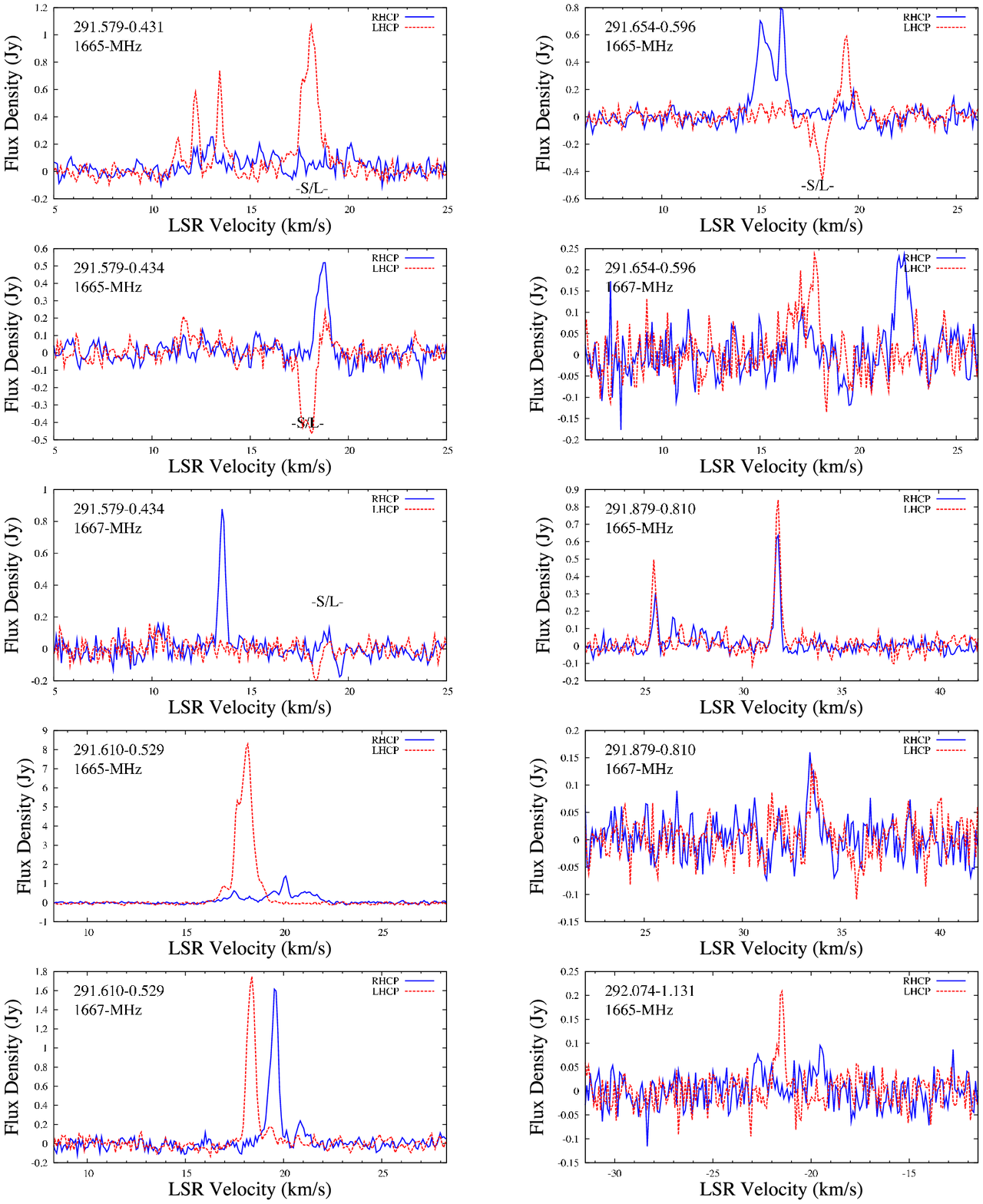}  
\caption{\small cont. The flux densities of 291.610$-$0.529 and 291.654$-$0.596 should be scaled by factors of 1.01 and 1.02 respectively to account for the primary beam correction. }
\label{RLspectra}
\end{center}
\end{figure*}

\begin{figure*}
\begin{center}
\addtocounter{figure}{-1}
\renewcommand{\baselinestretch}{1.1}
\includegraphics[width=18cm]{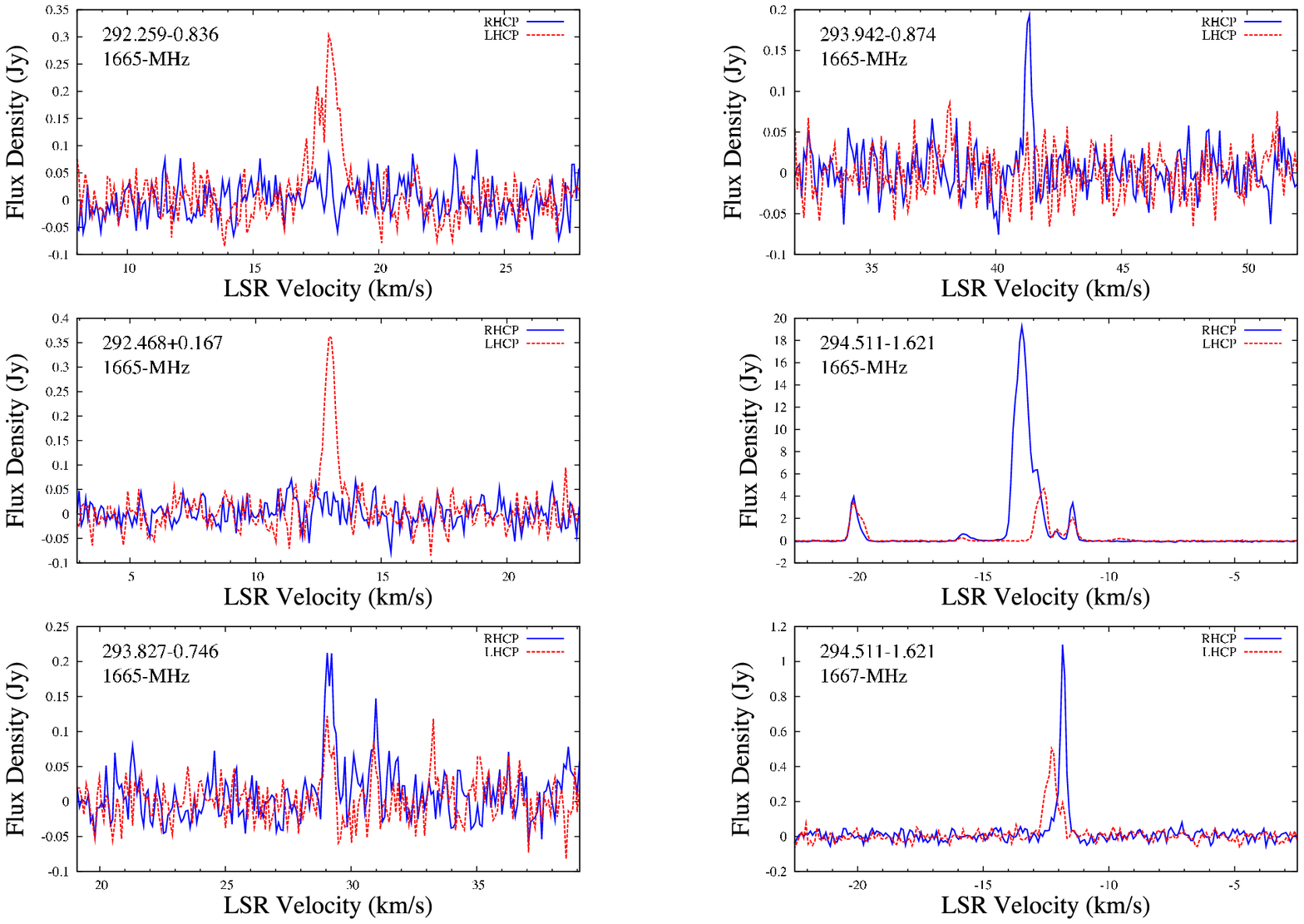}  
\caption{\small cont. The flux density of 292.259$-$0.836 should be scaled by a factor of 6.58 to account for the primary beam correction. }
\label{RLspectra}
\end{center}
\end{figure*}

\begin{table*} \centering \caption{\small Ground-state OH maser detections as measured with the ATCA. Positions are accurate to $\sim$0.4$''$. *denotes new sources. Previous source references are: CH87 - \citet{caswell87b}; C98 - \citet{caswell98}; B89 - \citet{braz89}; tLH96 - \citet{telintel96}; C04 - \citet{caswell04b}. $^{1}$Primary beam correction to be applied to listed flux density to account for offset position of source from primary beam. $^{2}$6.7-GHz methanol is classified as coincident if it lies within the positional errors, for those sites that are not coincident the offset is listed.} %check caswell04 is correct
\begin{tabular}{lcrcrlrrrll}
\hline
\multicolumn{1}{c}{Source Name} & \multicolumn{2}{c}{Equatorial Coordinates} & \multicolumn{1}{c}{Maser} & \multicolumn{2}{l}{Peak} & \multicolumn{1}{c}{Peak} & \multicolumn{2}{c}{Velocity} & \multicolumn{1}{l}{Coincident}& \multicolumn{1}{l}{Previous}\\
\multicolumn{1}{c}{(~~~$l$,~~~~~~~$b$~~~)}    &       RA(J2000)        &       Dec(J2000)       & Transition & \multicolumn{2}{l}{Flux~~~($\times$pbc$^{1}$)}& \multicolumn{1}{c}{Velocity}   & \multicolumn{2}{c}{Range}      &    6.7-GHz&Reference \\
\multicolumn{1}{c}{(~~~$^\circ$~~~~~~~$^\circ$~~~)} & (h~~m~~~s) & (~$^\circ$~~ $'$~~~~$''$) & \multicolumn{1}{c}{(MHz)} &  \multicolumn{2}{l}{(Jy)} & \multicolumn{1}{c}{(km\,s$^{-1}$ )}&   \multicolumn{2}{c}{(km\,s$^{-1}$ )}  &Methanol$^{2}$& \\
\hline
284.351$-$0.418&10 24 10.72& $-$57 52 33.8&1612&6.5& ($\times$1.00)&5.0&4.5&9.0&No ($\le$5$''$)& ({\it new transition})\\
&10 24 10.68& $-$57 52 33.9&1665&1.6 &($\times$1.00)&5.6&5.0&9.0&No ($\le$5$''$)&CH87, C98\\
&10 24 10.74& $-$57 52 34.2&1667&1.1& ($\times$1.00)&5.7&5.0&8.5&No ($\le$5$''$)&CH87, C98\\
284.694$-$0.361*&10 26 36.29& $-$58 00 34.5&1665&5.2& &13.0&11.5&15.0&Yes\\
285.263$-$0.050&10 31 29.87& $-$58 02 18.4&1665&54.0 & ($\times$1.08)&5.9&3.5&14.0 &No (5.3$'$)& CH87,  C98\\
&10 31 29.89& $-$58 02 19.2&1667&3.4 & ($\times$1.08)&3.7&2.0&10.0& No (5.3$'$) &CH87,  C98\\
287.371+0.644&10 48 04.41& $-$58 27 00.8&1665&1.7&&$-$4.0&$-$4.5&$-$3.0&Yes&tLH96,  C98\\
290.374+1.661&11 12 18.10& $-$58 46 21.3&1665&4.1&&$-$23.1&$-$24.0&$-$12.0&Yes&B89, C98\\
&11 12 18.09& $-$58 46 21.3&1667&1.8&&$-$22.6&$-$23.0&$-$12.0&Yes&B89, C98\\
291.274$-$0.709&11 11 53.36& $-$61 18 23.6 &1665&0.4&&$-$24.2&$-$25.0&$-$23.5&Yes&C04\\
291.579$-$0.431&11 15 05.81& $-$61 09 40.4&1665&0.8&&13.5&11.0&14.0& Yes&CH87,  C98\\
291.579$-$0.434*&11 15 05.23& $-$61 09 49.4&1665&0.7& ($\times$1.00)&18.8&18.0&19.5&No ($\le$10$''$)& \\
&11 15 05.20& $-$61 09 49.8&1667&0.9 &($\times$1.00)&13.6&13.0&14.0&No ($\le$10$''$)&\\
291.610$-$0.529&11 15 02.68& $-$61 15 48.9&1665&8.5& ($\times$1.01)&18.2&16.0&23.0&No (2.1$'$)& CH87,  C98\\
&11 15 02.67& $-$61 15 48.8&1667&1.7& ($\times$1.01)&18.4&18.0&22.0&No (2.1$'$)&CH87,  C98\\
291.654$-$0.596*&11 15 10.74& $-$61 20 32.3&1665&0.8& ($\times$1.02)&16.1&14.0&20.1&No (3.1$'$)&\\
&11 15 10.68& $-$61 20 33.5&1667&0.3& ($\times$1.02)&17.1&17.0&23.0&No (3.1$'$)&\\
291.879$-$0.810*&11 16 17.34& $-$61 37 21.0&1665&1.5&&31.8&25.0&32.5&Yes&\\
&11 16 17.30& $-$61 37 21.2&1667&0.3&&33.5&33.0&34.5&Yes&\\
292.074$-$1.131*&11 16 51.25& $-$61 59 32.3&1665&0.2&&$-$21.5&$-$23.0&$-$21.1&Yes&\\
292.259$-$0.836*& 11 19 12.58 & $-$61 46 54.9&1665&0.4& ($\times$6.58)&18.0&17.0&19.0&No (22.9$'$)&\\
292.468+0.167*&11 23 42.13& $-$60 54 33.9&1665&0.4&&12.9&12.0&14.0&Yes&\\
293.827$-$0.746*&11 32 05.63& $-$62 12 25.5&1665&0.3&&29.1&28.5&30.7&Yes&\\
293.942$-$0.874*&11 32 42.09& $-$62 21 47.8&1665&0.2&&41.3&40.8&41.8&Yes&\\
294.511$-$1.621&11 35 32.23& $-$63 14 42.8&1665&19.3&&$-$13.5&$-$20.9&$-$10.5&Yes&B89, C98\\
&11 35 32.25& $-$63 14 43.2&1667&1.3&&$-$11.9&$-$13.0&$-$11.0&Yes&B89, C98\\
\hline
\end{tabular} 
\label{resotable}
\end{table*}

\begin{table*}
\begin{minipage}{180mm}
\small
\centering
\caption{\small Flux densities of the Stokes parameters ($I$,$Q$,$U$,$V$) and linear polarization ($P$) for the ground-state OH features.
Stokes $I$, $Q$, $U$ and $V$ have an rms noise of $\sim$50 mJy. $P$ has an error of $\sim$70 mJy.  $p_{\rm l}$ is the fractional linear polarization and $p_{\rm c}$ is the fractional circular polarization. Where the linearly or circularly polarized flux density is below 5$\sigma$, upper limits are provided for the percentage polarization.
As the velocities are taken as the midpoint of channels there is an error of $\pm$0.5 channels (corresponding to $\sim$0.045 km s$^{-1}$).}
\begin{tabular}{c r r r r r r c c }
\\
\hline
\multicolumn{1}{l}{Feature}& \multicolumn{1}{r}{$V_{\rm LSR}$} & \multicolumn{1}{c}{$I$} & \multicolumn{1}{c}{$Q$} & \multicolumn{1}{c}{$U$} & \multicolumn{1}{c}{$V$} & \multicolumn{1}{c}{$P$} & \multicolumn{1}{c}{$p_{\rm l}$} & \multicolumn{1}{c}{$p_{\rm c}$}\\
\multicolumn{1}{l}{No.}   & (\,km\,s$^{-1}$) & (Jy) & (Jy)& (Jy)& (Jy)& (Jy)& \multicolumn{1}{c}{(\%)}& \multicolumn{1}{c}{(\%)}\\
\hline 
\multicolumn{2}{l}{284.351$-$0.418 - 1612-MHz}\\ 
1 & 4.99 & 6.49 & 0.74 & $-$1.20 & 6.02 & 1.41 & 22$\pm$1  & 93$\pm$1 \\ 
2 & 5.35 & 3.36 & 0.90 & $-$0.26 & $-$2.97 & 0.94 & 28$\pm$2  & 88$\pm$1 \\ 
3 & 6.17 & 0.69 & 0.09 & $-$0.06 & $-$0.35 & 0.11 & $\le$30  & 51$\pm$4 \\ 
4 & 6.72 & 2.03 & 0.19 & $-$0.08 & $-$1.67 & 0.21 &  $\le$15  & 82$\pm$1 \\ 
5 & 8.26 & 1.37 & $-$0.33 & $-$0.02 & $-$0.31 & 0.33 &  $\le$30  & 23$\pm$3 \\ 
6 & 8.35 & 1.37 & $-$0.04 & $-$0.09 & $-$1.00 & 0.10 &   $\le$15  & 73$\pm$1 \\ 
\multicolumn{2}{l}{284.351$-$0.418 - 1665-MHz}\\ 
1 & 5.62 & 1.63 & $-$0.08 & 0.20 & 0.58 & 0.22 & $\le$20  & 35$\pm$4 \\ 
2 & 5.89 & 0.72 & $-$0.16 & $-$0.06 & $-$0.34 & 0.17 & $\le$35  & 47$\pm$4 \\ 
3 & 6.24 & 0.51 & $-$0.06 & $-$0.12 & 0.06 & 0.14 & $\le$45  & $\le$25 \\ 
4 & 7.47 & 0.58 & $-$0.12 & $-$0.00 & 0.51 & 0.12 & $\le$35  & 88$\pm$16 \\ 
5 & 7.65 & 0.43 & $-$0.11 & $-$0.12 & 0.49 & 0.16 & $\le$60  & 115$\pm$25 \\ 
\multicolumn{2}{l}{284.351$-$0.418 - 1667-MHz}\\  
1 & 5.74 & 1.11 & $-$0.02 & 0.03 & $-$0.31 & 0.03 &  $\le$10 & 28$\pm$3 \\ 
2 & 6.10 & 0.57 & $-$0.04 & 0.08 & $-$0.10 & 0.09 & $\le$30  & $\le$25 \\ 
3 & 7.68 & 0.36 & 0.07 & 0.06 & 0.14 & 0.09 & $\le$60  & $\le$60 \\ 
4 & 7.94 & 0.29 & $-$0.03 & $-$0.17 & 0.15 & 0.17 & $\le$100  & $\le$80 \\ 
\multicolumn{2}{l}{284.694$-$0.361 - 1665-MHz}\\ 
1 & 12.58 & 1.15 & $-$0.16 & 0.16 & 0.14 & 0.22 & $\le$25  & $\le$20 \\ 
2 & 13.02 & 5.16 & $-$1.64 & 0.39 & 3.37 & 1.68 & 33$\pm$2  & 65$\pm$2 \\ 
3 & 14.16 & 0.19 & $-$0.01 & $-$0.11 & 0.12 & 0.11 & $\le$100  &$\le$100 \\ 
\multicolumn{2}{l}{285.263$-$0.050 - 1665-MHz}\\ 
1 & 2.75 & 2.59 & 0.10 & 0.04 & $-$2.55 & 0.11 &  $\le$10  & 99$\pm$1 \\ 
2 & 4.25 & 12.85 & 0.06 & $-$0.14 & 12.46 & 0.15 &  $\le$10  & 97$\pm$1 \\ 
3 & 5.92 & 54.02 & $-$0.22 & $-$0.69 & $-$52.94 & 0.73 &  $\le$10  & 98$\pm$1 \\ 
4 & 7.24 & 1.80 & $-$0.02 & 0.06 & $-$1.64 & 0.06 &  $\le$10  & 91$\pm$1 \\ 
5 & 7.77 & 1.23 & $-$0.01 & $-$0.02 & $-$1.13 & 0.02 &  $\le$10 & 93$\pm$1 \\ 
6 & 8.83 & 3.48 & 0.05 & $-$0.14 & 3.50 & 0.14 &  $\le$10  & 101$\pm$3 \\ 
7 & 9.71 & 4.17 & $-$0.03 & $-$0.08 & 4.35 & 0.08 &  $\le$10  & 104$\pm$4 \\ 
8 & 10.85 & 0.27 & 0.02 & 0.02 & 0.29 & 0.03 & $\le$40  & 106$\pm$40 \\ 
9 & 11.29 & 0.47 & $-$0.01 & 0.02 & 0.51 & 0.03 &  $\le$20 & 108$\pm$22 \\ 
10 & 12.17 & 8.29 & $-$0.03 & $-$0.02 & 8.54 & 0.04 &  $\le$10  & 103$\pm$3 \\ 
\multicolumn{2}{l}{285.263$-$0.050 - 1667-MHz}\\ 
1 & 2.52 & 1.23 & 0.06 & 0.04 & $-$1.32 & 0.07 &  $\le$15  & 108$\pm$8 \\ 
2 & 3.05 & 0.66 & $-$0.03 & 0.08 & $-$0.65 & 0.08 & $\le$25  & 98$\pm$1 \\ 
3 & 3.67 & 3.35 & 0.22 & $-$0.07 & 3.28 & 0.23 &  $\le$10  & 98$\pm$3 \\ 
4 & 6.92 & 0.75 & $-$0.02 & $-$0.09 & $-$0.13 & 0.09 & $\le$25  & $\le$25 \\ 
5 & 8.86 & 0.42 & 0.01 & 0.03 & $-$0.25 & 0.03 & $\le$25  & 59$\pm$5 \\ 
\multicolumn{2}{l}{287.371+0.644 - 1665-MHz}\\ 
1 & $-$3.96 & 1.74 & $-$0.08 & $-$0.04 & $-$1.77 & 0.09 &  $\le$10  & 101$\pm$1 \\ 
2 & $-$3.69 & 1.33 & 0.03 & $-$0.03 & $-$1.32 & 0.04 & $\le$10  & 99$\pm$1 \\ 
\multicolumn{2}{l}{290.374+1.661 - 1665-MHz}\\ 
1 & $-$23.15 & 4.07 & $-$0.34 & 0.29 & $-$4.03 & 0.45 & $\le$15  & 99$\pm$1 \\ 
2 & $-$19.46 & 0.34 & $-$0.20 & 0.03 & $-$0.12 & 0.20 & $\le$90  & $\le$50\\ 
3 & $-$17.17 & 0.20 & $-$0.11 & $-$0.01 & $-$0.17 & 0.11 & $\le$100 & $\le$90 \\ 
4 & $-$16.73 & 0.23 & $-$0.00 & $-$0.01 & $-$0.14 & 0.01 &  $\le$40  & $\le$70 \\ 
5 & $-$12.42 & 0.26 & $-$0.05 & $-$0.03 & $-$0.27 & 0.06 & $\le$55  & 102$\pm$2 \\ 
\multicolumn{2}{l}{290.374+1.661 - 1667-MHz}\\ 
1 & $-$22.59 & 1.82 & $-$0.25 & 0.09 & $-$1.68 & 0.26 & $\le$20  & 93$\pm$1 \\ 
2 & $-$13.00 & 0.22 & 0.06 & 0.03 & $-$0.22 & 0.06 & $\le$70  & $\le$100 \\ 
\multicolumn{2}{l}{291.274$-$0.709 - 1665-MHz}\\ 
1 & $-$24.20 & 0.45 & $-$0.01 & 0.08 & 0.30 & 0.08 & $\le$36  & 68$\pm$19 \\ 
2 & $-$24.64 & 0.25 & 0.07 & 0.06 & 0.13 & 0.09 & $\le$73  & $\le$82 \\ 
\multicolumn{2}{l}{291.579$-$0.431 - 1665-MHz}\\ 
1 & 13.45 & 0.82 & -0.02 & -0.02 & -0.66 & 0.03 &  $\le$15  & 80$\pm$1 \\ 
\multicolumn{2}{l}{291.579$-$0.434 - 1665-MHz}\\ 
1 & 18.81 & 0.75 & $-$0.02 & 0.02 & 0.29 & 0.03 &  $\le$15  & 38$\pm$9 \\ 
\multicolumn{2}{l}{291.579$-$0.434 - 1667-MHz}\\ 
1 & 13.57 & 0.90 & 0.18 & 0.16 & 0.85 & 0.24 & $\le$35  & 95$\pm$11 \\ 
\end{tabular} 
\label{poltable}
\end{minipage}
\end{table*}

\begin{table*}
\addtocounter{table}{-1}
\begin{minipage}{180mm}
\small
\centering
\caption{\small cont}
\begin{tabular}{c r r r r r r c c }
\\
\hline
\multicolumn{1}{l}{Feature}& \multicolumn{1}{r}{$V_{\rm LSR}$} & \multicolumn{1}{c}{$I$} & \multicolumn{1}{c}{$Q$} & \multicolumn{1}{c}{$U$} & \multicolumn{1}{c}{$V$} & \multicolumn{1}{c}{$P$}  & \multicolumn{1}{c}{$p_{\rm l}$} & \multicolumn{1}{c}{$p_{\rm c}$} \\
\multicolumn{1}{l}{No.}   & (\,km\,s$^{-1}$) & (Jy) & (Jy)& (Jy)& (Jy)& (Jy)& \multicolumn{1}{c}{(\%)}& \multicolumn{1}{c}{(\%)} \\
\hline 
\multicolumn{2}{l}{291.610$-$0.529 - 1665-MHz}\\ 
1 & 16.94 & 1.11 & $-$0.06 & $-$0.02 & $-$0.65 & 0.06 &  $\le$15  & 58$\pm$2 \\ 
2 & 17.65 & 5.74 & $-$0.02 & $-$0.14 & $-$4.93 & 0.14 &  $\le$10  & 86$\pm$1 \\ 
3 & 18.18 & 8.50 & $-$0.10 & $-$0.09 & $-$8.06 & 0.13 &  $\le$10  & 95$\pm$1 \\ 
4 & 19.50 & 0.68 & 0.00 & $-$0.06 & 0.52 & 0.06 &  $\le$20  & 77$\pm$13 \\ 
5 & 20.11 & 1.42 & 0.05 & 0.04 & 1.35 & 0.06 &  $\le$10  & 95$\pm$7 \\ 
6 & 21.26 & 0.52 & $-$0.07 & $-$0.07 & 0.58 & 0.10 & $\le$35 & 113$\pm$21 \\ 
\multicolumn{2}{l}{291.610$-$0.529 - 1667-MHz}\\ 
1 & 18.39 & 1.73 & 0.06 & $-$0.02 & $-$1.76 & 0.06 &  $\le$10  & 102$\pm$2 \\ 
2 & 19.53 & 1.70 & $-$0.03 & $-$0.01 & 1.53 & 0.03 &  $\le$10  & 90$\pm$6 \\ 
3 & 20.76 & 0.28 & 0.08 & $-$0.00 & 0.11 & 0.08 & $\le$60  & $\le$70 \\ 
\multicolumn{2}{l}{291.654$-$0.596 - 1665-MHz}\\ 
1 & 15.01 & 0.80 & $-$0.06 & 0.01 & 0.60 & 0.06 &  $\le$20 & 75$\pm$11 \\ 
2 & 16.06 & 0.85 & $-$0.08 & $-$0.07 & 0.74 & 0.11 & $\le$20  & 86$\pm$11 \\ 
3 & 19.41 & 0.66 & 0.11 & $-$0.06 & $-$0.52 & 0.12 & $\le$30  & 79$\pm$2 \\ 
\multicolumn{2}{l}{291.654$-$0.596 - 1667-MHz}\\ 
1 & 17.07 & 0.28 & $-$0.05 & 0.04 & $-$0.12 & 0.06 & $\le$55  & $\le$55 \\ 
2 & 17.77 & 0.25 & $-$0.06 & 0.01 & $-$0.22 & 0.06 & $\le$60  & $\le$95 \\ 
3 & 22.35 & 0.23 & 0.02 & 0.07 & 0.24 & 0.07 & $\le$70  & $\le$100 \\ 
\multicolumn{2}{l}{291.879$-$0.810 - 1665-MHz}\\ 
1 & 25.49 & 0.68 & 0.18 & 0.59 & $-$0.31 & 0.62 & 90$\pm$17  & 45$\pm$4 \\ 
2 & 31.82 & 1.48 & $-$0.00 & $-$1.40 & $-$0.20 & 1.40 & 95$\pm$8  & 14$\pm$3 \\ 
\multicolumn{2}{l}{291.879$-$0.810 - 1667-MHz}\\ 
1 & 33.53 & 0.26 & $-$0.05 & 0.09 & $-$0.02 & 0.10 & $\le$75  &  $\le$30 \\ 
\multicolumn{2}{l}{292.074$-$1.131 - 1665-MHz}\\ 
1 & $-$21.48 & 0.22 & $-$0.01 & $-$0.01 & $-$0.20 & 0.02 &  $\le$45  & 89$\pm$3 \\ 
\multicolumn{2}{l}{292.259$-$0.836 - 1665-MHz}\\ 
1 & 17.57 & 0.23 & $-$0.05 & 0.11 & $-$0.19 & 0.12 & $\le$100  & $\le$85\\ 
2 & 18.01 & 0.39 & $-$0.05 & 0.09 & $-$0.22 & 0.11 & $\le$60 & $\le$65 \\ 
\multicolumn{2}{l}{292.468+0.167 - 1665-MHz}\\ 
1 & 12.92 & 0.36 & 0.12 & $-$0.07 & $-$0.36 & 0.14 & $\le$65  & 100$\pm$1 \\ 
\multicolumn{2}{l}{293.827$-$0.746 - 1665-MHz}\\ 
1 & 29.05 & 0.33 & 0.23 & $-$0.00 & 0.09 & 0.23 & $\le$100  & $\le$50\\ 
2 & 30.99 & 0.19 & 0.08 & $-$0.20 & 0.10 & 0.21 & $\le$100  & $\le$100 \\ 
\multicolumn{2}{l}{293.942$-$0.874 - 1665-MHz}\\ 
1 & 41.25 & 0.23 & $-$0.06 & $-$0.00 & 0.13 & 0.06 & $\le$65  & $\le$90 \\ 
\multicolumn{2}{l}{294.511$-$1.621 - 1665-MHz}\\ 
1 & $-$20.15 & 7.45 & 6.56 & 2.65 & 0.46 & 7.08 & 95$\pm$2  &  6$\pm$1 \\ 
2 & $-$15.84 & 0.87 & $-$0.12 & $-$0.24 & 0.33 & 0.27 & $\le$40 & 38$\pm$8 \\ 
3 & $-$13.46 & 19.27 & $-$0.06 & $-$0.25 & 19.22 & 0.26 &  $\le$10  & 100$\pm$1 \\ 
4 & $-$12.85 & 9.75 & $-$0.50 & $-$2.12 & 2.98 & 2.18 & 22$\pm$1  & 31$\pm$1 \\ 
5 & $-$12.06 & 1.82 & 1.50 & $-$0.21 & $-$0.22 & 1.51 & 83$\pm$6  & $\le$15 \\ 
6 & $-$11.44 & 5.45 & $-$0.20 & $-$2.56 & 1.33 & 2.57 & 47$\pm$2  & 24$\pm$1 \\ 
\multicolumn{2}{l}{294.511$-$1.621 - 1667-MHz}\\ 
1 & $-$12.29 & 0.61 & 0.10 & $-$0.04 & $-$0.40 & 0.11 & $\le$25  & 65$\pm$3 \\ 
2 & $-$11.85 & 1.29 & 0.24 & 0.21 & 0.91 & 0.32 & $\le$35  & 70$\pm$7 \\ 
\end{tabular} 
\label{poltable}
\end{minipage}
\end{table*}

\subsection{Detected masers not associated with star formation}
\begin{figure}
\begin{center}
\renewcommand{\baselinestretch}{1.1}
\includegraphics[width=8.2cm]{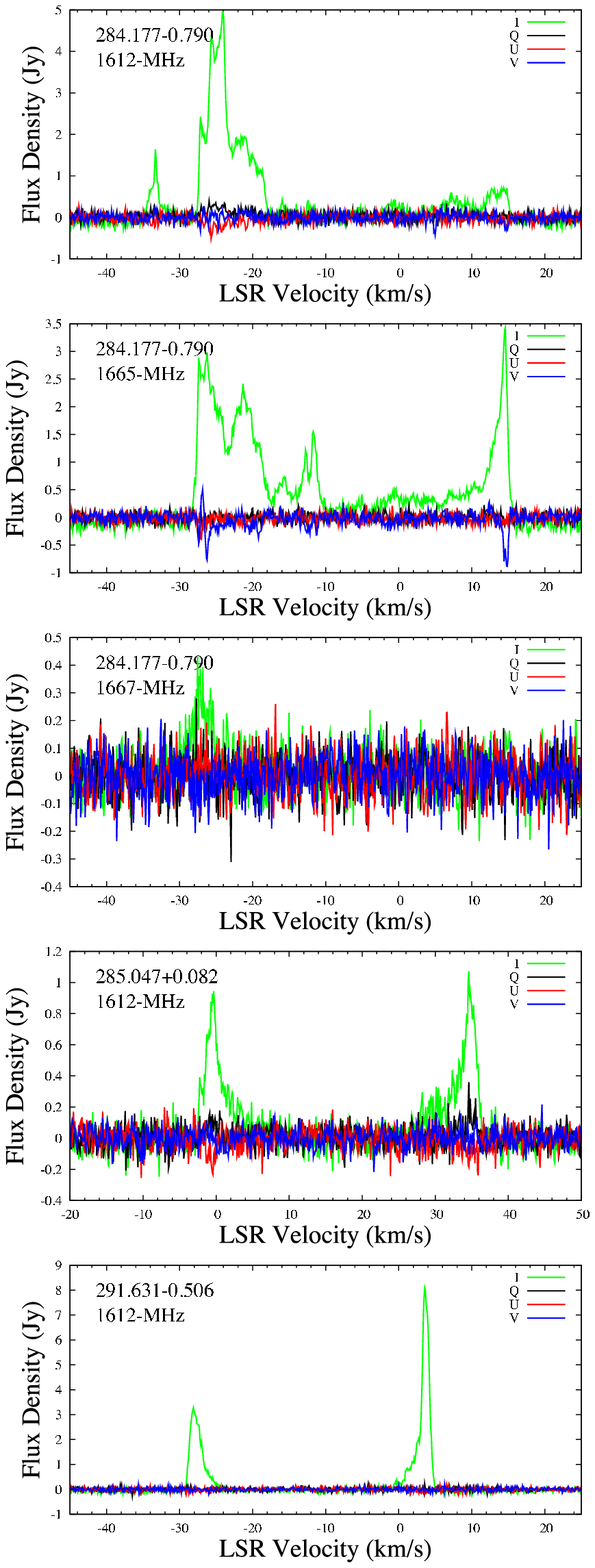}  
\caption{\small Three sources detected in the survey which are not associated with star formation. The first, 284.177$-$0.790, is seen at 1612, 1665 and 1667 MHz. The other two are seen at 1612 MHz only. All are offset from the pointing centre, requiring primary beam corrections of approximately 10, 2.5 and 1.1 to be applied to the plotted flux density values.}
\label{OTHERSspectra}
\end{center}
\end{figure}

We serendipitously detected three sites of OH maser emission not associated with star formation (284.177$-$0.790, 285.047+0.082 and 291.631$-$0.506). Spectra for all these sites are shown in Figure\,\ref{OTHERSspectra}, but were found significantly offset from the pointing centre (25$'$, 18$'$ and 5$'$ respectively) and are subject to primary beam corrections. Two of these sites (285.047+0.082 and 291.631$-$0.506) exhibit a clear `double horn' spectral profile at the 1612-MHz transition, characteristic of late-type stars.  They have been previously observed \citep[e.g.][]{telintel91}, but we greatly improve their positional accuracy.  

285.047+0.082 was discovered by  \citet{caswell81b}, with a positional uncertainty of 35$''$, whereas our new position of RA(J2000) 10$^{\rm h}$30$^{\rm m}$36.517$^{\rm s}$, Dec(J2000) $-$57$^{\circ}$48$'$52.22$''$ has a positional uncertainty of 0.4$''$.  Our position confirms the association with an infrared counterpart, to its measured positional accuracy of 7$''$ \citep{epchtein82}, and with IRAS 10287$-$5733 suggested by \citet{telintel91}.  Our present flux density values, after correction for the primary beam (see Figure\,\ref{OTHERSspectra}) are comparable to the archival measurements.

291.631$-$0.506 corresponds with the source listed at the IRAS position 11123$-$6101 (291.563$-$0.593) by \citet{telintel91}, whose 1612-MHz measurements with the Parkes telescope were made 1986 May. Our improved position for the maser of  RA(J2000) 11$^{\rm h}$15$^{\rm m}$16.495$^{\rm s}$, Dec(J2000) $-$61$^{\circ}$15$'$01.28$''$ (rms error 0.4$''$) is offset from the IRAS source by more than 6.6$'$, showing that the IRAS identification is incorrect and revealing that the 1986 May observations were taken near the half-power point of the Parkes beam, at 1612 MHz; consequently, the te Lintel Hekkert flux densities must be underestimated by a factor of two.  For comparison, our measured flux densities are weaker by about 30\% than the corrected 1986 values.  For this maser, the difference probably demonstrates real variability, as commonly occurs in the late-type stars.

The third source, 284.177$-$0.790, was detected at the 1612-, 1665- and 1667-MHz transitions, with a wide velocity range, but less clear spectral profile. This was previously positioned by \citet{caswell98} and our position is consistent. Our flux densities, after primary beam correction, are within the range of archival values, which are variable.

\begin{table*}
\begin{minipage}{180mm}
\small
\centering
\caption{\small Zeeman pairs associated with the Carina-Sagittarius spiral arm tangent. The splitting factors used are: 0.123\,km\,s$^{-1}$ mG$^{-1}$ for 1612 MHz; 0.590\,km\,s$^{-1}$ mG$^{-1}$ for 1665 MHz; 0.354\,km\,s$^{-1}$ mG$^{-1}$ for 1667 MHz; 0.113\,km\,s$^{-1}$ mG$^{-1}$ for 1720 MHz; and 0.0564\,km\,s$^{-1}$ mG$^{-1}$ for 6035 MHz (see main text). References are: CV95 - \citet{caswell95c}, C04 - \citet{caswell04a}. Qualities are defined as: A = unambiguous  Zeeman pair associated to within the positional errors of 0.4$''$; and B = where there are more than one RHCP or LHCP component that can be associated. Magnetic field errors are based on the spectral accuracy of the components (a product of an error of half the velocity resolution on each component). Field strengths from the current observations which are in italics are below 5$\sigma$ significance. Distances are kinematic based on the velocity of the midpoint of the RHCP and LHCP components and are defined the same as those in Table\,\ref{targetstable}.}
\begin{tabular}{l r r r c c c c c l}
\\
\hline
Source& Freq& \multicolumn{1}{r}{RHCP $V_{\rm LSR}$} & \multicolumn{1}{r}{LHCP $V_{\rm LSR}$}& \multicolumn{1}{c}{RHCP $S_{p}$} & \multicolumn{1}{c}{LHCP $S_{p}$} & \multicolumn{1}{c}{$B$}  &Quality & Distance&reference\\
& (MHz)& (\,km\,s$^{-1}$) & (\,km\,s$^{-1}$)  & (Jy)& (Jy)& (mG) & &(kpc)& \\
\hline 
284.351$-$0.418 	&1612  & 4.99 & 5.35 & 6.26& 3.17 & $-$2.9$\pm$0.5& B & 5.2$\pm$0.8&(this work)  \\
                                	&1612  & 8.17 & 8.35 & 1.03& 1.18 & {\it $-$1.5$\pm$0.5}& B & 5.2$\pm$0.8&(this work)    \\
                            		&1665  & 5.36 & 5.71 & 1.26& 0.55 & $-$0.6$\pm$0.1& B & 5.2$\pm$0.8&(this work)   \\
                            		&1667  & 5.83 & 5.74 & 0.41& 0.71 & {\it +0.3$\pm$0.1}& B  & 5.2$\pm$0.8&(this work)   \\
284.694$-$0.361 & 1665 & 13.02 & 12.93 &4.27&0.92&{\it +0.2$\pm$0.1}&B&6.0$\pm$0.7&(this work)\\
285.263$-$0.050 & 1665 & 4.25 & 2.75 & 12.65 & 2.57 & +2.5$\pm$0.1& A &  5.2$\pm$0.8& (this work) \\
& 1667 & 3.67 & 2.52 & 3.32 & 1.27 & +3.2$\pm$0.1 & A & 5.2$\pm$0.8& (this work) \\
& 6035 & 9.10 & 8.60 & 3.10 & 3.00 & +8.9$\pm$1.3& A &  5.8$\pm$0.7& CV95 \\ 
290.375+1.666 &1720 & $-$20.00& $-$20.90 & 0.35 & 0.58 & +8.0$\pm$4.4 & A&  2.9$\pm$1.9& C04 \\ 
291.610$-$0.529 &1665 & 20.11&18.18& 1.39&8.28&+3.3$\pm$0.1&B& 8.2$\pm$0.6& (this work) \\
 &1667 & 19.53 & 18.39 & 1.62 & 1.74 & +3.2$\pm$0.1& A & 8.2$\pm$0.6&(this work) \\
294.511$-$1.621 & 1665 & $-$11.44 & $-$12.58 & 3.39 & 4.68 & +1.9$\pm$0.1& B &6.1$\pm$0.7 & (this work) \\
 & 1667 & $-$11.85 & $-$12.29 & 1.10 & 0.50 & +1.2$\pm$0.1& A &6.1$\pm$0.7 &(this work) \\
  & 6035 &$-$12.00 & $-$12.07& 4.60&4.60 & {\it +1.2$\pm$1.3} &A&6.1$\pm$0.7 &CV95\\ 
\end{tabular} 
\label{RLtable}
\end{minipage}
\end{table*}

\section{Discussion}
\begin{figure}
\begin{center}
\renewcommand{\baselinestretch}{1.1}
\includegraphics[width=8cm]{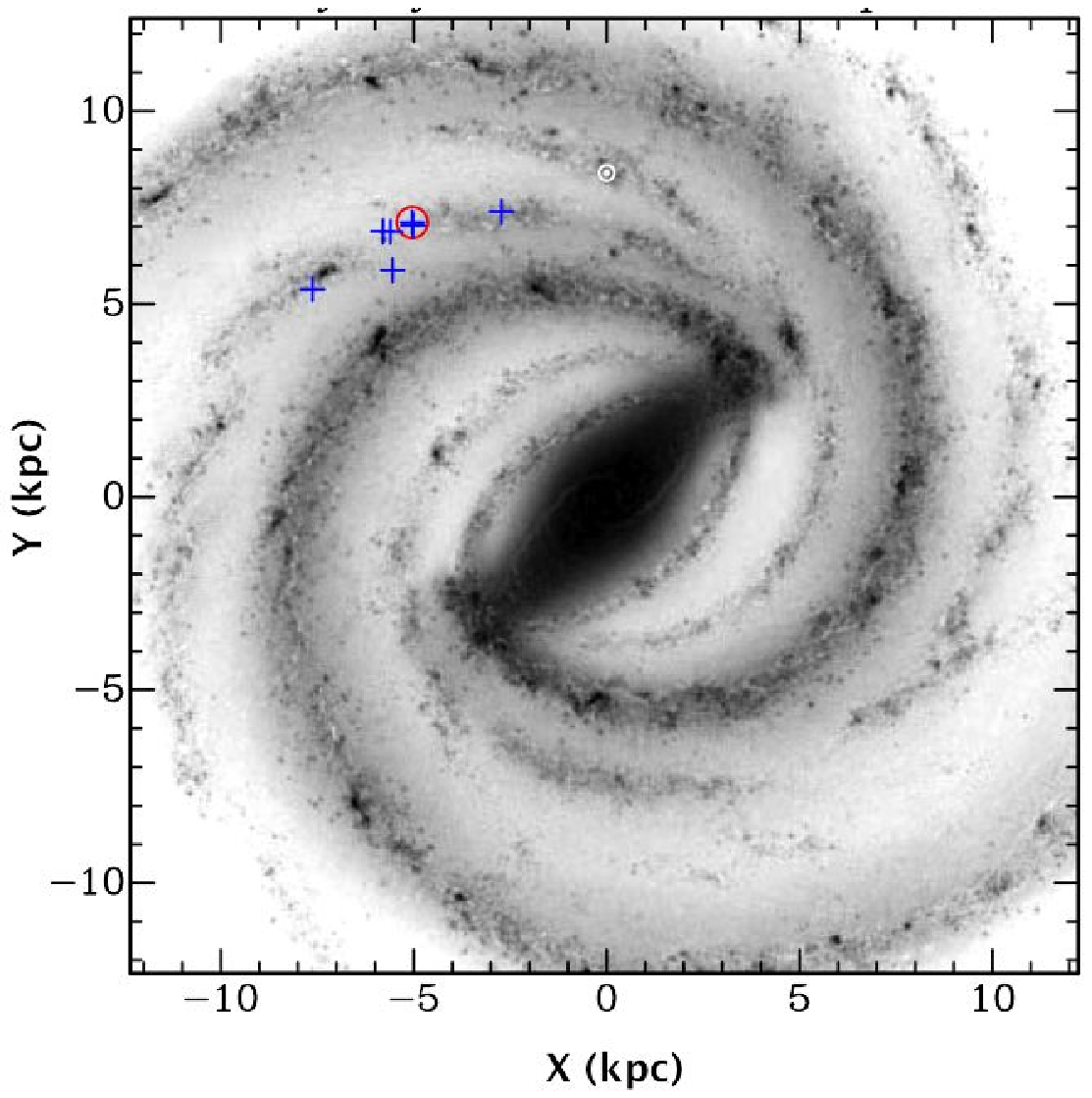}  
\caption{\small  Magnetic field direction (blue pluses positive, red circles negative) as inferred from the current Zeeman splitting measurements overlaid on the informed artist impression of the Milky Way (R. Hurt: NASA/JPL-Caltech/SSC). This shows how the magnetic field measurements relate to Carina-Sagittarius spiral arm. We see that the fields are consistent in direction with the exception of the low significance measurements towards 284.351$-$0.418. The magnetic field direction is discussed in full in section \ref{importantfield}, but under the current conventions would be directed away from the sun for positive measurements and towards for negative. Galactic rotation is clockwise in this figure. The sun is located at (0,8.4).}
\label{topdownplot}
\end{center}
\end{figure}

Here we consider the implications of the magnetic field measurements presented in section \ref{resultssection} (the 11 measurements of the current study combined with the three previous measurements collectively spread across six sites of high-mass star formation). For measurements where we find multiple transitions for the same star-forming site, only 284.351$-$0.418 has an inconsistent measurement, that of the 1667-MHz transition but this is not a statistically significant result ($\le$3$\sigma$). It should also be noted that for sites such as this with multiple transitions, higher resolution observations reveal the transitions to be spatially offset, and thus could sample minor differences in the field (as has been shown in VLBI observations, such as those by \citealt{caswell11vlbi1}). Considering only the seven `A' class measurements (spread across four sites of high-mass star formation) we find all Zeeman splitting measurements  are positive with magnetic field strengths of 1.2 to 8.9 mG. Considering only the ten field measurements with statistical significance above 5$\sigma$ we find eight positive Zeeman measurements with the same range of field strengths. We hence see a strong indication for a coherent field orientation in the tangent of the Carina-Sagittarius arm which, under the currently accepted convention (see Appendix), is directed away from us, in a direction counter-clockwise to Galactic rotation (Figure\,\ref{topdownplot}). We first discuss the significance of the field coherence and then discuss the importance of the field orientation.

\subsection{Coherent field orientation within the Carina-Sagittarius spiral arm tangent}
The presence of a coherent orientation of magnetic field direction has important implications for the fields permeating the regions of high-mass star formation traced by the masers. Our first consideration is the statistical significance of the predominant magnetic field orientation that we see. If the magnetic field vectors were pointed randomly within a three dimensional sphere, for any given field there would be an equal chance of a positive or negative Zeeman splitting measurement. Of the 14 Zeeman pair measurements the most likely result would be of the order of 5 to 9 positive field measurements (81\% statistical probability). We see 11 positive and three negative, which has a 2\% statistical probability of occurring by chance. Considering just the `A' quality measurements of Table\,\ref{RLtable}, of which there are seven, we find all are positive, which has  $<$1\% statistical probability of occurring by chance. There is hence a strong indication that these measurements indicate a large-scale coherently ordered field orientation in the Carina-Sagittarius arm. Using the distances listed in Table\,\ref{RLtable} we estimate the physical scale over which magnetic field coherence is seen: the nearest positive magnetic field is located at 2.9$\pm$1.9 kpc (290.375+1.666) and the furthest at 8.2$\pm$0.6 kpc (291.610$-$0.529); giving a physical separation of 5.3$\pm$2.0 kpc. 

More generally, this implies (as per discussion of \citealt{han07}) that large-scale magnetic fields either play a dominant role in the development of molecular clouds and the process of star formation \citep[in accord with the recent review by][]{crutcher03}, or the fields are unaffected by the process. Either way, they are conserved during the contraction to the small scales and high densities of star formation despite the dynamically disruptive processes involved. Theoretical modelling of high-mass star formation concurs, forecasting that the direction of the magnetic field pervading the larger molecular cloud will be conserved on the collapse of material into star forming regions \citep[e.g.][]{li96,allen03}.

Coherently oriented magnetic field directions across regions of star formation have been found before, with \citet{davies74} providing the first suggestion, finding coherent magnetic field directions across eight sites of OH maser emission (based on a simple pattern of clockwise Galactic rotation). \citet{reid90} then found coherence over scales of a few kiloparsecs in a sample of 17 OH magnetic field measurements spread across the Galactic plane (although with a pattern more complex than Davies suggested, with the field orientations consistent within spiral arms, rather than with a global clockwise direction). \citet{fish03b} found consistent magnetic field orientations from six masers in the second and third Galactic quadrants (which are free from kinematic distance ambiguities) and from six masers with near kinematic distances $\le$2 kpc in the first and fourth quadrants. The inner and outer Galaxy masers showed opposite field directions, nominally in agreement with rotation measure estimates \citep[e.g.][]{brown07,vaneck11} but there were disagreements elsewhere, such as the masers in the fourth quadrant on the far side of the Galactic centre.
\citet{fish03b} also found 80\% coherence of field orientation in the Norma arm (increased to 100\% recently with the allocation of the discrepant sources to the 3-kpc arms by \citealt{green09b}). Most recently, \citet{han07} compiled previous (heterogeneous) OH maser observations that provide hints of large-scale patches of field coherence elsewhere in the Galaxy.
 
\subsection{The Galactic magnetic field direction within the Carina-Sagittarius spiral arm tangent}\label{importantfield} 
We compare our OH Zeeman measurements with large-scale Galactic magnetic field orientations inferred from Faraday rotation measurements of pulsars \citep[][]{taylor93b,han99,han06} and extragalactic sources \citep{brown07}. Faraday rotation of linearly polarized radiation causes a frequency dependent rotation of the measured polarization position angle, $\phi$, defined by the IAU starting from North and increasing through East.  Rotation measures probe the product of the magnetic field, $\mathbf{B}~({\rm \mu G})$, and the electron density, $n_e~{\rm {(cm^{-3})}}$, along the line of sight, $\mathbf{r}$ (pc), such that the measured linear polarization angle of a source emitting at a position angle $\chi_0$ is
\begin{equation}
\chi =\chi_0 + 0.81 \lambda^2 \, \int_{\rm source}^{\rm observer} n_{\rm e} \, \mathbf{B} \cdot d{\bf r}  = \chi_0 + {\rm RM} \lambda^2
\label{eq:RM}
\end{equation}
where RM is the rotation measure in units of rad\,m$^{-2}$ (see \citealt{brentjens05}).
A positive rotation measure arises from a magnetic field towards the observer, and the convention implicit in equation \ref{eq:RM} is that a magnetic field towards the observer is denoted by a positive sign. Between Galactic longitudes 280$^{\circ}$ and 295$^{\circ}$ collectively the rotation measure is overwhelmingly positive, indicating a high degree of magnetic field coherence in this region of the Galaxy and a field directed towards us. 

\begin{figure*}
\begin{center}
\renewcommand{\baselinestretch}{1.1}
\includegraphics[width=14cm]{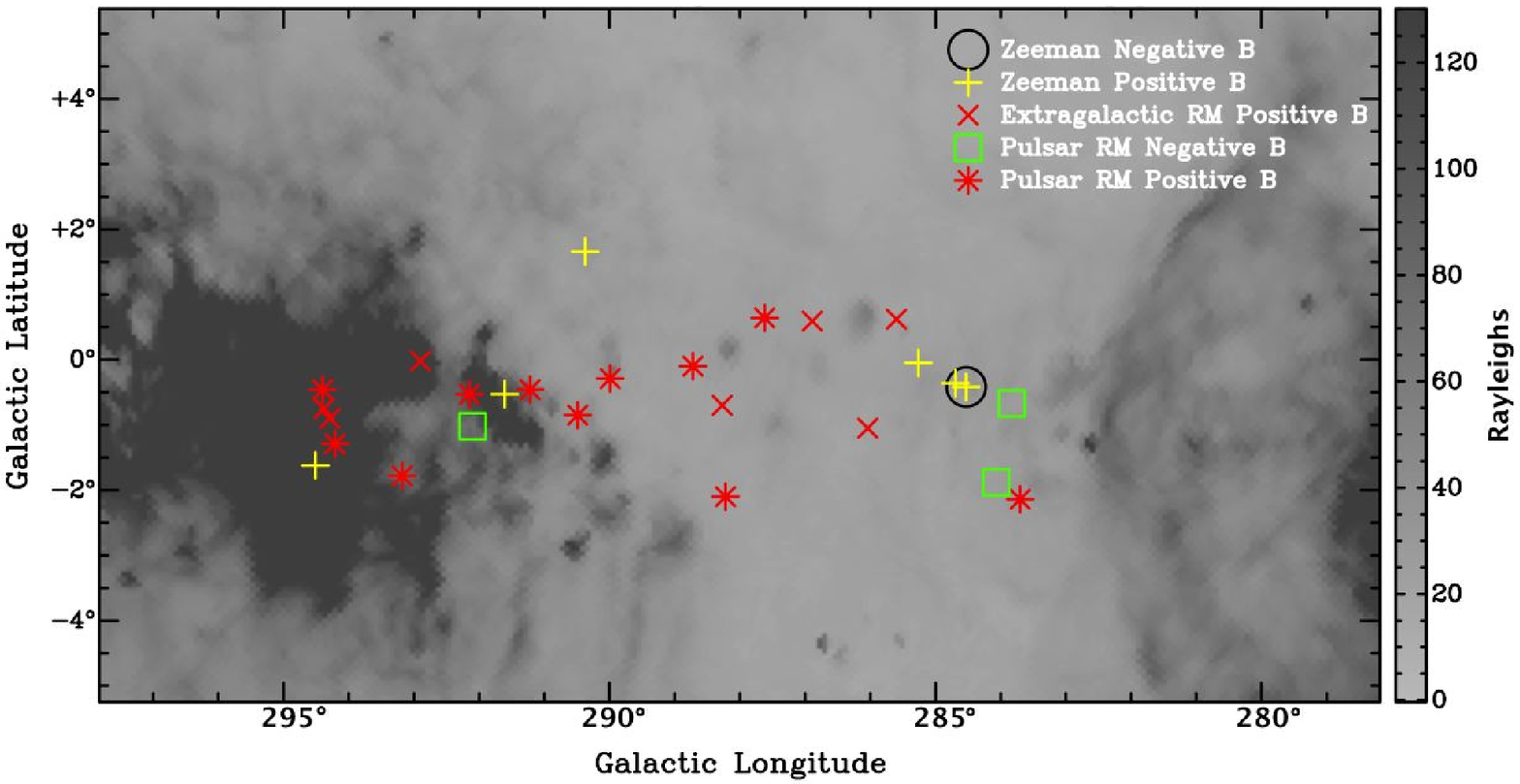}  
\caption{\small Comparison of magnetic field directions from rotation measures and Zeeman splitting measurements overlaid on the composite H$\alpha$ emission of \citet{finkbeiner03} using the Southern H-Alpha Sky Survey Atlas data of \citet{gaustad01}. The H$\alpha$ emission indicates regions of ionized hydrogen. Yellow pluses are positive $B$ field Zeeman measurements (conventionally fields away from us, but see discussion in section \ref{importantfield}) and a black circle is the sole site of negative $B$ field Zeeman measurements (a field towards us); red asterisks are positive rotation measures (fields towards us) from pulsars; green squares are negative rotation measures (fields away from us) from pulsars \citep{taylor93b,han99,han06} and red crosses are positive rotation measures (fields towards us) from extragalactic sources \citep{brown07}. The rotation measures, from both pulsars and extragalactic sources, are all along lines of sight in the proximity of, or directly passing through, H{\sc ii} regions \citep[][references therein]{nota10}.}
\label{zeemanRMcomp}
\end{center}
\end{figure*}

Figure\,\ref{zeemanRMcomp} compares the field orientations of our Zeeman measurements with those from rotation measures of pulsars (\citealt{han99,han06} and \citealt{taylor93b}) and extragalactic sources (\citealt{brown07}, derived from the Southern Galactic Plane Survey: \citealt{mcclure05,haverkorn06}). In the figure, we single out the 21 rotation measures for lines of sight believed to be passing through known H{\sc ii} regions (those rejected by  \citealt{nota10}), of which 18 are positive. 14 are towards pulsars (11 positive, three negative), with the pulsars estimated to lie at distances between 4 and 17 kpc \citep[][references therein]{han06} and seven measurements (all positive) are towards extragalactic sources (clearly beyond 17 kpc). We choose lines of sight through H{\sc ii} regions (all of which lie within the Carina-Sagittarius arm)  as there are indications that the rotation measures for these lines of sight will be influenced by the field experienced within the H{\sc ii} region, due to the significantly enhanced electron density present \citep[e.g.][]{harvey11}. These measurements thus provide a good comparison with the fields from the Zeeman measurements which represent the in situ magnetic fields of sites of star formation within the Carina-Sagittarius arm. The pulsar rotation measures could be singled out as uniquely measuring purely the integrated Galactic magnetic field.  However, from the compilation of \citet{oppermann12}, it is clear that, within this longitude region, extragalactic and pulsar rotation measures (with sightlines through H{\sc ii} or not) indicate the same dominant field orientation. We also note that 13 distinct lines of sight spread over the 15 arcmin extent of supernova remnant G292.2$-$0.5 (within the tangent of the Carina-Sagittarius arm, at a distance of 8.4 kpc) are also all large and positive \citep{caswell04c}. As can be seen in Figure\,\ref{zeemanRMcomp}, the measurements from Zeeman splitting and the rotation measures indicate oppositely oriented coherent fields, a result also suggested in the work of \citet[][and references therein]{han07}, but based on fewer Zeeman measurements (three within the longitude range). 

It is theoretically possible that the rotation measures could, instead of sampling the Carina-Sagittarius arm, be biased by a field permeating an over-density of electrons elsewhere in the Galaxy: for the rotation measures between Galactic longitudes 280$^{\circ}$ and 295$^{\circ}$ (two thirds of which are pulsars) this would be located either in the local arm or in the distant Perseus arm. The Perseus arm, which is not tangential at these longitudes and lies at a distance in excess of 13\,kpc, is unlikely to significantly affect the rotation measure, with only three of the pulsars having distance estimates comparable to the Perseus arm (J1112$-$6103, J1119$-$6127, J1133$-$6250). A local H{\sc ii} region \citep[see for example][]{harvey11}, could affect the field orientation, but no local large (of the order of 10 square degrees) H{\sc ii} region is known to exist at these longitudes (see Figure\,\ref{zeemanRMcomp}). Additionally, the study of \citet{kronberg11}, which shows the smooth symmetry of the rotation measures across Galactic longitudes, would argue against these longitudes being biased by a local object with a different field orientation. 

The magnetic field measurements in this region therefore imply that the techniques of Zeeman splitting and rotation measures indicate opposite field orientations.\footnote{The choice of designating a positive or negative value for a magnetic field towards the observer, historically differs between astronomers interpreting Faraday rotation and those interpreting Zeeman splitting \citep{manchester72,verschuur69}. However, this is merely a result of differing conventions and if the two techniques were to trace the same magnetic field, when expressed as fields towards or away from the observer, the techniques should agree.  The apparent disagreement suggested by our measurements prompts us to re-examine (in the Appendix) the relation between magnetic field direction and the sense of Zeeman splitting.} 
Further comparison (once the full survey data set is available) of magnetic field orientation across the Galaxy (with larger statistics) will reveal whether the different techniques yield consistently opposite field orientations, or if this part of the Galaxy is unusual.

\section{Summary} 
We present the preliminary observations of a large-scale project to observe the magnetic fields found in regions of high-mass star formation traced by 6.7-GHz methanol masers. We observed 23 sites of methanol maser emission between the Galactic longitude range 280$^{\circ}$ to 295$^{\circ}$. The detection rate towards sites of 6.7-GHz methanol maser emission was approximately 50\% (11 of the 23 sites).  We detected a total of 17 sites of OH maser emission, nine sites new discoveries of the survey. We found predominantly 1665-MHz emission, with only one site of 1612- and no 1720-MHz emission detected across any of the sites. Fractional linear polarization ranged from 22 to 95\% and fractional circular polarization from 6 to 100\%. Combined with previous measurements, 14 Zeeman pairs were found spread across six sites of high-mass star formation, with all measurements above 5$\sigma$ yielding the same magnetic field direction. These measurements are spread across a physical scale of 5.3$\pm$2.5\,kpc and thus indicate that a coherent magnetic field is experienced by the sites of high-mass star formation within the Carina-Sagittarius arm.  Under the current convention, this field direction is opposite to that determined from rotation measures. The detection rate and consistency of the direction of the magnetic fields of this preliminary study is very encouraging for the full Galactic survey.

\section*{Acknowledgments}
The authors thank W.H.T.~Vlemmings, M.D.~Gray and D.H.F.M.~Schnitzeler for useful discussions and the staff of the Australia Telescope Compact Array. The Australia Telescope Compact Array is part of the Australia Telescope which is funded by the Commonwealth of Australia for operation as a National Facility managed by CSIRO.\\

\bibliographystyle{mn2e} \bibliography{UberRef}

\appendix

\section[]{Field direction convention in Zeeman splitting measurements}
\begin{figure}
\begin{center}
\renewcommand{\baselinestretch}{1.1}
\includegraphics[width=8cm]{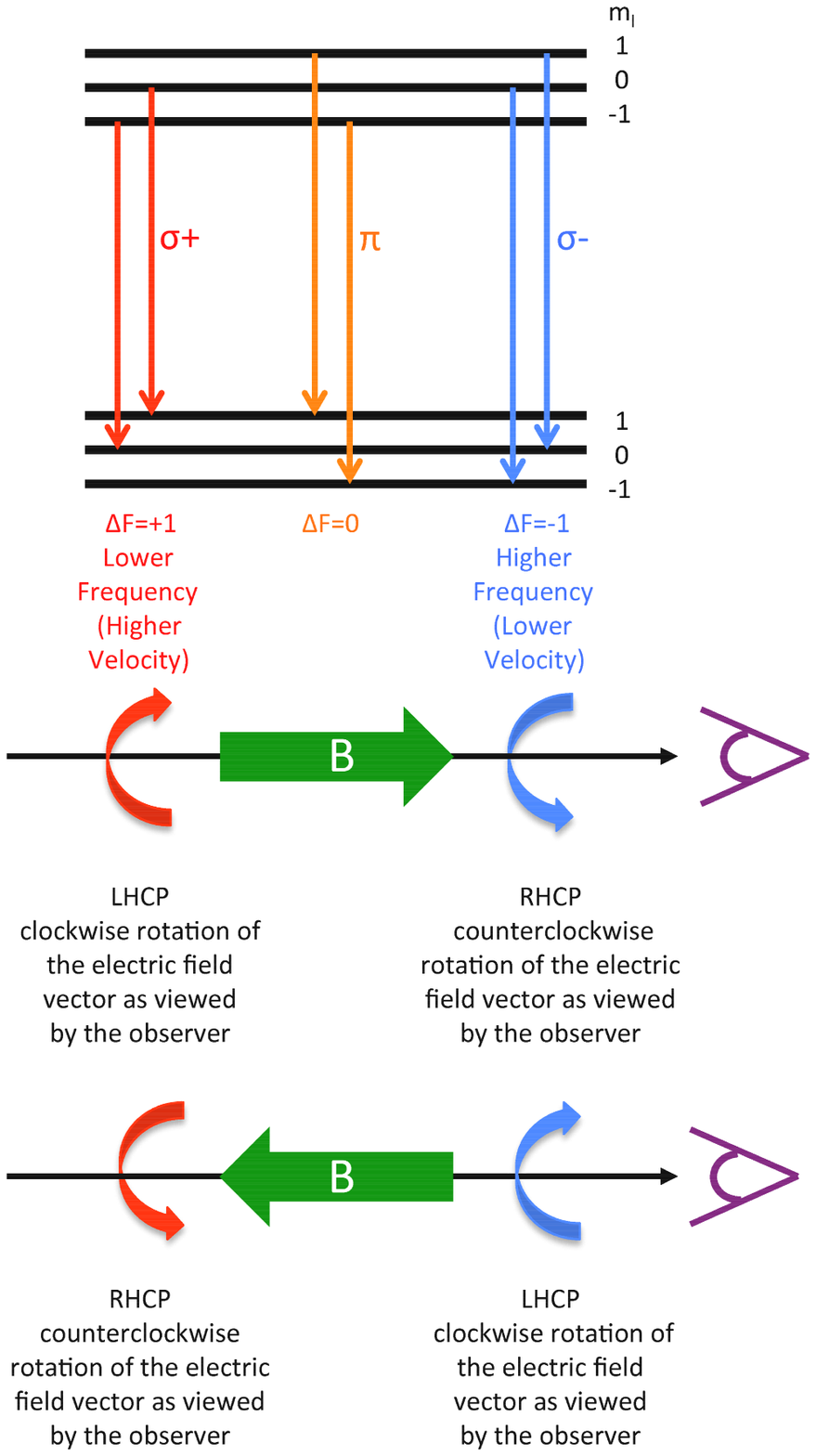}  
\caption{\small Summary of Zeeman splitting conventions for the example of the 1665-MHz maser transition ($^{2}\Pi_{3/2}$, $J = 3/2$). In this figure F is the total angular momentum and $m_l$ is the magnetic quantum number of the orbital subshell. The $\sigma$ components are the elliptically polarized components and $\pi$ the linearly polarized component. The black arrow head represents the direction of emission, towards the location of the observer (purple symbol) at the right of the diagram. The RHCP and LHCP are defined according to the IEEE standard.}
\label{zeemanconventions}
\end{center}
\end{figure}

\begin{figure}
\begin{center}
\renewcommand{\baselinestretch}{1.1}
\includegraphics[width=8cm]{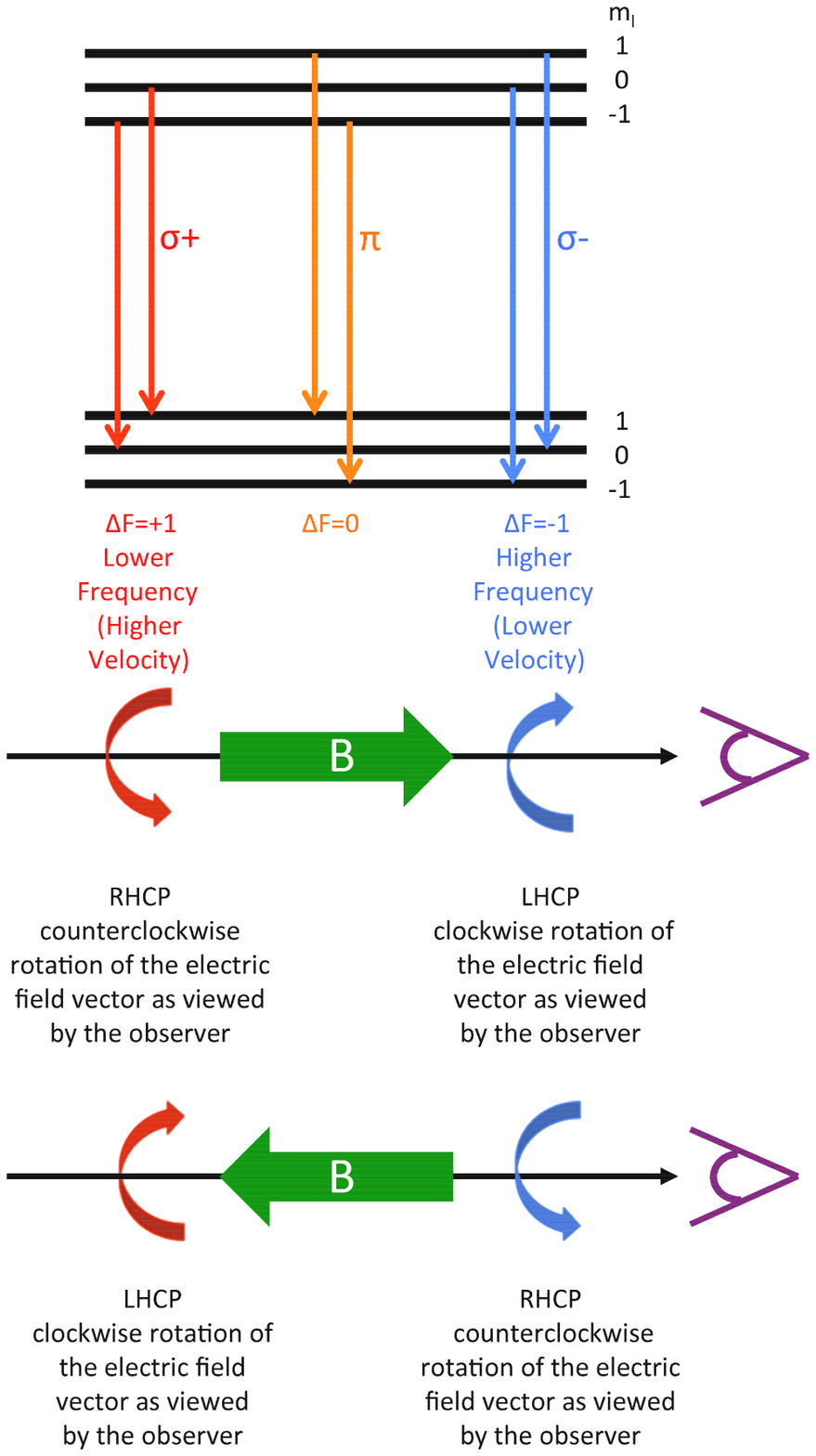}  
\caption{\small An alternative interpretation of Zeeman splitting conventions, which might be inferred from the description by \citet{cook75,cook77}, for the example of the 1665-MHz maser transition ($^{2}\Pi_{3/2}$, $J = 3/2$). In this figure F is the total angular momentum and $m_l$ is the magnetic quantum number of the orbital subshell. The $\sigma$ components are the elliptically polarized components and $\pi$ the linearly polarized component. The black arrow head represents the direction of emission, towards the location of the observer (purple symbol) at the right of the diagram. The RHCP and LHCP are defined according to the IEEE standard.}
\label{ZeemanSplittingCook75}
\end{center}
\end{figure}

In the Zeeman measurements presented here we have adopted the IEEE definition for RHCP and LHCP and the IAU convention for Stokes $V$. Accordingly, RHCP and LHCP are defined by the IEEE standard such that an observer would view RHCP as counterclockwise circular polarization (i.e. rotation of the electric field vector in a counterclockwise direction as the wave travels from the source to the observer along the line of sight). Stokes $V$ is defined as the IAU standard, RHCP minus LHCP (opposite  to that adopted by the pulsar community). Additionally it has become common practice for maser Zeeman splitting measurements to take a positive magnetic field value as representing a field oriented away from the observer. Figure \ref{zeemanconventions} is a schematic summary of the Zeeman splitting and the conventions. Unfortunately there is an ambiguity in tying the Zeeman splitting components (positive and negative $\sigma$) to the RHCP and LHCP emission which is often overcome with an arbitrary assignment. The approach described above, which is now the commonly accepted convention, is such that where the RHCP component of a Zeeman pair is at higher velocity, the magnetic field is interpreted as away from the observer. However,  the literature reveals a history of different interpretations \citep[discussed extensively in][]{hamaker96,robishaw08} and it is often difficult to find a clear description of which $\sigma$ component corresponds to which handedness of polarization. An early example within the maser community for potential ambiguity is that of \citet{cook75,cook77}, which describes the direction of the velocity shift of masing gas as being in the same direction as the direction of positive magnetic field. One interpretation of this is that for a (positive) magnetic field directed away from the observer, the Zeeman splitting causes the positive $\sigma$ component to shift in velocity in the direction away from the observer (to a lower frequency, higher local standard of rest velocity), and that this will be observed as LHCP emission. This is shown schematically in Figure\,\ref{ZeemanSplittingCook75}, which demonstrates (in comparison with Figure \ref{zeemanconventions}) the impact on observed RHCP and LHCP components for a given magnetic field direction (leading to a convention opposite to that implied by Figure \ref{zeemanconventions}).

\clearpage

\label{lastpage}

\end{document}